\documentclass[aps,pre,a4paper,10pt,twocolumn]{revtex4-1}

\usepackage{graphicx}
\usepackage{subfigure}
\usepackage{float}
\usepackage{bm}        
\usepackage{amssymb}   
\usepackage{amsmath}
\usepackage[utf8]{inputenc}
\usepackage[T1]{fontenc}
\usepackage{lmodern}
\usepackage{afterpage}

\usepackage{color}
\usepackage{braket}

\def\braket#1{\mathinner{\langle{#1}\rangle}}

\newcommand{\ba}[1]{\overline{#1}}

\newcommand{\de}{\partial}

\begin{document}
\title{Cold denaturation of RNA secondary structures with loop entropy and quenched disorder}

\author{Flavio Iannelli}
\email{iannelli.flavio@gmail.com} 
\affiliation{Humboldt-Universit\"at zu Berlin, Institut f\"ur Physik, Newtonstra{\ss}e 15, 12481 Berlin}

\author{Yevgeni Mamasakhlisov}
\affiliation{Department of Molecular Physics, Yerevan State University, 1 Alex Manougian Street, Yerevan 0025, Armenia}

\author{Roland R. Netz}
\affiliation{Fachbereich Physik, Freie Universit\"at Berlin, 14195 Berlin, Germany}

\begin{abstract}
We study the folding of RNA secondary structures with quenched sequence randomness by means of the constrained annealing method. A thermodynamic phase transition is induced by including the conformational weight of loop structures. In addition to the expected melting at high temperature, a cold melting transition appears.
Our results suggest that the cold denaturation of RNA found experimentally is in fact a continuous phase transition triggered by quenched sequence disorder. We calulate both hot and cold melting critical temperatures for the competing energy scenario between favorable and unfavorable base pairs and present  a phase diagram as a function of  the loop exponent and temperature.
\end{abstract}

\maketitle

\section{Introduction}
\label{intro}
Ribonucleic acids are biopolymers that are crucial to all living systems, they process and transmit genetical  informations and take part in many important cellular activities \cite{gesteland}. The RNA primary structure is a chain sequence which consists of four bases U, A, G and C, while the secondary structure is the listing of base pairings occurring in the more complex tertiary structure. 
An interesting  phenomena emerging at low temperatures is the folding of these molecules into their native conformations.
Like in the case of proteins, the folding of RNA structures is crucial for  understanding their biological functions and  has been vastly studied, grounded mostly in the idea of hierarchical folding \cite{tinoco}.
In this scheme the primary structure alone determines the folding mechanism and therefore the secondary structure forms independently of the tertiary structure. Such a  great simplification makes it possible to exactly compute the secondary structure partition function of given RNA sequences. 
The molecule folding can be qualitatively and quantitatively addressed via the experimentally observable helicity degree defined as the average fraction of paired bases, which increases when lowering the temperature in the standard scenario. 
Here we will focus on a two-letter alphabet binding energy model with symbols chosen from the subset  $\{$U,A$\}$, in the spirit of \cite{higgs,roland1,bund,ricci,bund2} which physically corresponds to the hydrophilic-hydrophobic model for protein folding \cite{dill}. 
Although this approximation is an oversimplification, it has proven to reproduce in a reasonable fashion the thermodynamics of these molecules and it is able to capture the essential physics behind the folding as well as the glass transition \cite{ricci}.
The so-called Watson-Crick base pairs UA, and GC,  are the most stable hydrogen bonds.
Since each base is essentially planar and its conformations are limited, every RNA secondary structure is defined by a list of pairings $(i,j)$ with each position appearing at most once.
In addition for any two base pairings $(i,j)$ and $(k,l)$, we only consider nested base pairings, where $i<k<l<j$, and independent base pairings, where $i<j<k<l$.
A third possibility would be to include pseudoknots, which are rare in real RNA, where one has $i<k<j<l$, but they can be excluded as a first approximation \cite{tinoco,nussinov}. 
This choice defines a hierarchical structure for the RNA conformation so that a recursive equation for the partition function can be used. 
Furthermore, since the exclusion of pseudoknots  discards the configurations that are not planar, when considering disordered sequences the system features frustration \cite{par,ricci}.
When the molecule is not in the native state, i.e. when there is a high number of unpaired bases, also loops play a crucial role and it is not rare that these are plentiful even at low temperatures.
The configurational entropy contribution $\Delta S_{n}^{loop} \sim k_B \ln n^{-c}$ for loops of length $n$ is characterized by the universal loop exponent $c$ \cite{degennes}. For ideal polymers, which are modeled as simple random walks, the exponent is  $c_{RW}=D/2$ where $D$ is the spatial dimensionality. 
Instead, for self-avoiding walks one finds in three dimensions $c_{SAW} \approx 1.76$, which  even increases further in real polymers since $c$ depends also on the number of helical strands emerging from loops \cite{duplantier}.
For the model with no disorder, homopolymeric RNA, in the range $2<c \lesssim  2.479$ a phase transition from the folded to the unfolded state, usually referred to as the molten phase \cite{bund}, is known to occur  when the temperature increases up to the melting point $T_m(c)$ \cite{kafri} and the $c$-dependent critical exponents have been analytically obtained \cite{einert1}.
A finite loop exponent has been shown to significantly improve salt-dependent RNA folding compared with experiments \cite{einert3}.

In this paper we address the influence of a loop exponent $c\ne0$ on the behavior of RNA secondary structures with random sequences where the disorder is quenched, i.e. fixed. This allows us to characterize a generic RNA molecule and in fact most of the results obtained here apply also to DNA. 
A very interesting phenomenon that  characterizes these polymers and proteins is the \emph{cold denaturation} \cite{pace,privalov0,privalov,calda}.
The denaturation of proteins and polymers rising the temperature is a consequence of the increase in configurational entropy. Denaturation when lowering the temperature is usually interpreted in terms of hydrophobic interactions. Experimentally, denaturation can be inferred by the presence of peaks in the specific heat which physically follow from an abrupt increase of the system entropy \cite{mick1,mick2,finkel,sedlmeier}. 
In this paper we give an alternative explanation for cold denaturation  in terms of quenched disorder which itself weakens the secondary structure formation at low temperature.
The double peak behavior of $C_V$ turns out to be associated with two different melting temperatures  of the RNA secondary structure.

\begin{figure*}
\centering
  \includegraphics[scale=0.2]{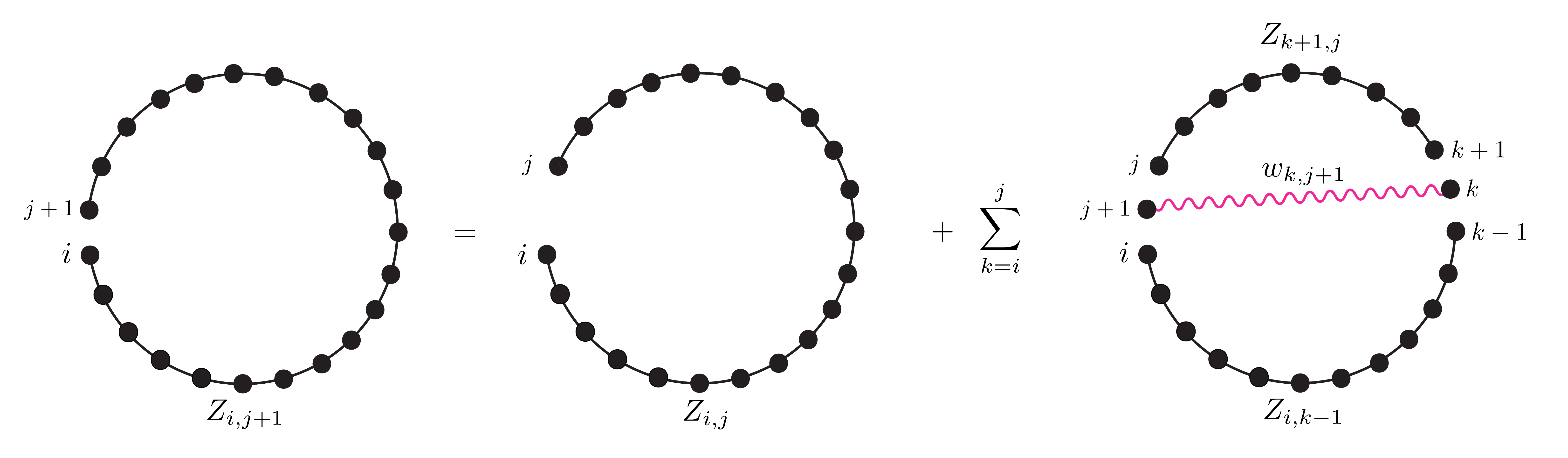}
  \caption{(Color online) Hierarchical recursive  scheme for the partition function eq. (\ref{recupart}). The subchain partition function from base $i$ to base $j+1$ is the sum of the partition function from base $i$ to base $j$ and the partition functions related to all nested or independent pairings formed with base $j+1$ by any base $k\in(i,j)$.  }
  \label{hartree}
\end{figure*} 

\section{Disordered RNA}
\subsection{The model}

For a sequence $h$ of length $N$ we define the base pairing matrix $\mathcal{S}$, which completely determines the secondary structure, as the $N\times N$ symmetric matrix with components $s_{i,j}$ equal to unity if $(i,j)$ are paired and vanishing otherwise. 
If stack energies are neglected, the Hamiltonian for a given sequence configuration can be written as a sum over non-repeated base pairs
\begin{eqnarray}
\mathcal{H} (\mathcal{S},h) = \sum_{ (i,j) \in \mathcal{S}} \epsilon_{i,j} =  \sum_{ 1\le i < j \le N} s_{i,j}\epsilon_{i,j}.
\label{hamiltonian}
\end{eqnarray}
Here we take the simplest non trivial pairing energy function as the sum of a constant and a disorder term in the spirit of \cite{bund2} as
\begin{eqnarray}
\epsilon_{i,j} = \epsilon_0 + \epsilon h_i h_j,
\label{epsilonij}
\end{eqnarray}
which can be easily generalized to a four-letter alphabet RNA.
The sign of the constant $\epsilon_0$ defines the nature of the background interaction between nucleotides. If $\epsilon_0>0$ the interaction is repulsive and attractive otherwise. The second term is the product of two independent variables of the form of a spin glass model for neural networks \cite{amit}, multiplied by a constant energy $\epsilon$. We assign Ising variables to each base along the chain so that $h_{i} = +1$ if $i$ is the nucleotide U and $h_{i} = -1$ if it is A. Contrary to the base pairing matrix elements $s_{i,j}$, which are free to evolve within the dynamics of the system, the site sequence variables $h_{i}$ are frozen and not free to rearrange to minimize the total energy. 
Having fixed the sign of the background interaction $\epsilon_0$,  the absolute value of the ratio $\epsilon / \epsilon_0$ is the only relevant parameter characterizing the system behavior, which is a direct consequence of the two-state model adopted here.  
For each sequence $h$ we take $h_i$ as quenched random independent and identically distributed variables so that the probability distribution factorizes as 
\begin{eqnarray}
\mathcal{P}(h) = \prod_{i=1}^N \rho (h_i).
\label{probdistr}
\end{eqnarray}
This construction, which is analytically more manageable, is supported by the fact that no strong correlations are found in the base type occurrence \cite{ricci}. Defining the probability of finding the base U as $p\equiv \rho(h_i = +1)$,  the probability distribution factors can be written as 
\begin{eqnarray}
\rho (h_i) = p\delta(h_i - 1) + (1-p)\delta(h_i+1).
\label{probabilita}
\end{eqnarray}
Due to symmetry one only needs to explore the parameter range $0<p<0.5$.

\subsection{Partition function}

The partition function of a given sequence is the sum over all allowed realizations of the base paring matrix 
\begin{eqnarray}
Z_N ( h ) = \sum_{\{\mathcal{S}\}} e^{-\beta \mathcal{H}(\mathcal{S},h)},
\label{partfun}
\end{eqnarray}
where $\beta=(k_B T)^{-1}$ and $\{\mathcal{S}\}$ denotes the set of all secondary structures without pseudoknots  for the given sequence $h$.
The free energy is obtained by performing the quenched average,  denoted by $\ba{(\cdots)}$, of the disordered free energy 
\begin{eqnarray}
f(h)  = -\frac{1}{ \beta N} \ln Z_N(h),
\end{eqnarray} 
over the disorder distribution eq. \eqref{probdistr}, yielding 
\begin{eqnarray}
 \ba{f(h)} = \sum_{\{h\}}  \mathcal{P}(h) f(h) = -\frac{1}{\beta  N} \ba{\ln Z_N(h)}.
\label{fque}
\end{eqnarray} 
For sufficiently large chains the physical properties of the system do not depend on the specific disorder realisation $\{h\}$ and the free energy self-averages \cite{par}
\begin{eqnarray}
\lim_{N\rightarrow \infty} f(h) = \ba{f(h)}.
\end{eqnarray} 
Numerically the partition function is usually obtained via the recursive equation of the restricted partition function \cite{mcca}  
\begin{eqnarray}
Z_{i,j+1} = Z_{i,j} + \sum_{k = i}^{j}w_{k,j+1}   Z_{i,k-1} Z_{k+1,j},
\label{recupart}
\end{eqnarray}
as illustrated in Fig. \ref{hartree}. 
On the right-hand side the first term corresponds to the probability that base $j+1$ is not paired, and the summation term corresponds to the probability associated to all possible nested or independent pairings given that position $k$ forms a paring with position $j+1$ with statistical weight 
\begin{eqnarray}
w_{k,j+1} \equiv \exp\left(-\beta \epsilon_{k,j+1}\right).
\end{eqnarray}
This recursive equation allows to compute the exact partition function $Z_N = Z_{1,N}$ without pseudoknots in a time of order $\mathcal{O}(N^3)$, starting with the boundary conditions  $Z_{i,i} = Z_{i,i-1} = 1$, $\forall i$.

\begin{figure}
\centering
  \includegraphics[scale=0.5]{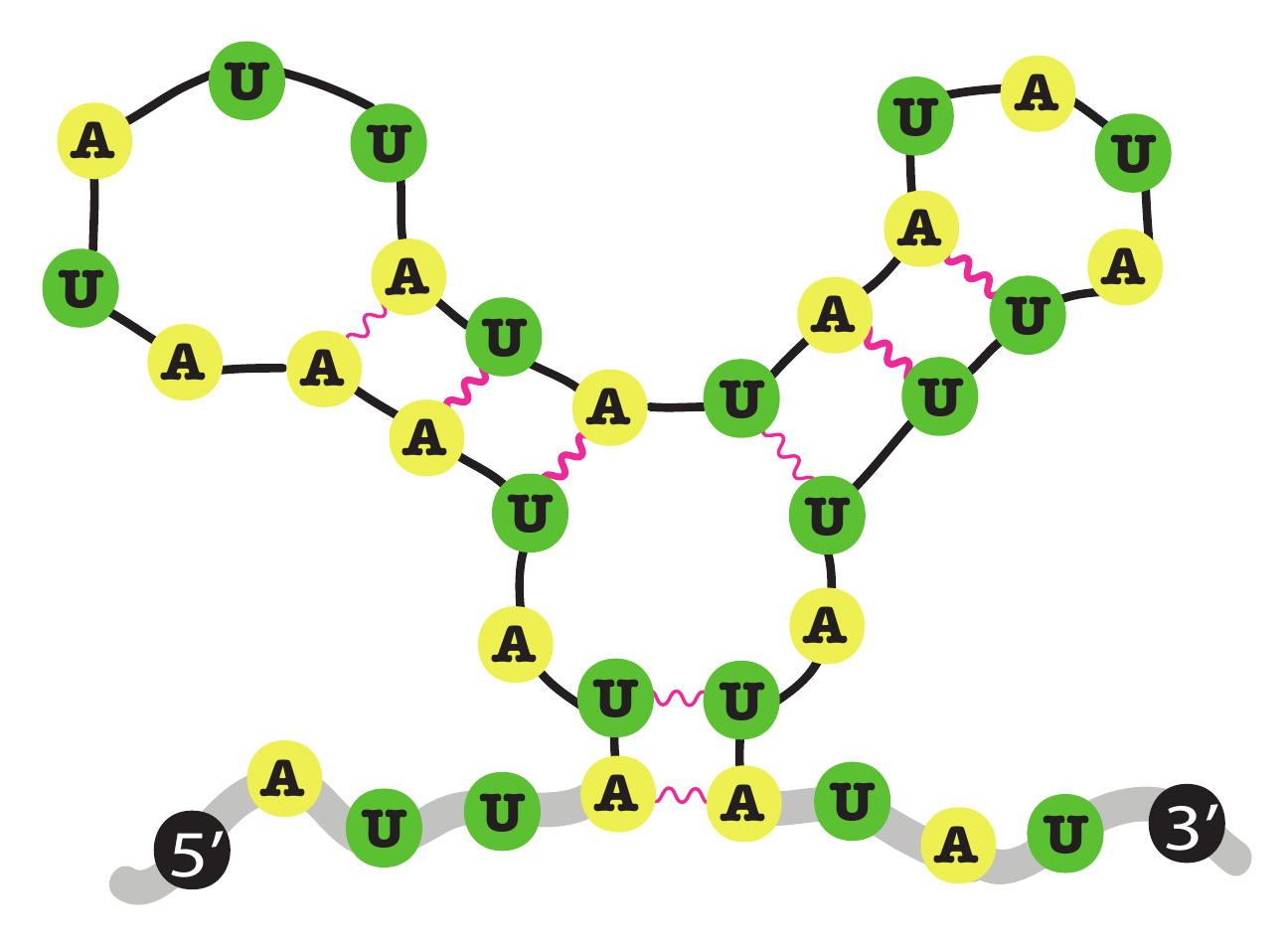}
  \caption{(Color online) Secondary structure representation with stacks and loops. Nucleotides are the green and yellow dots for U and A respectively. The wavy red lines identify the hydrogen bonds for favorable (thicker line) and unfavorable (thinner line) base pairs and the black solid lines the nested backbone links. In thick grey are the non-nested backbone links which are accounted by the upper index $M$ in the partition function eq. (\ref{secext}). For this structure going from the $5'-$ to the $3'-$end the number of total links is $N=34$ out of which $M=8$ are the non-nested.}
  \label{backbone}
\end{figure} 

\begin{figure*}
\centering
  \includegraphics[scale=0.2]{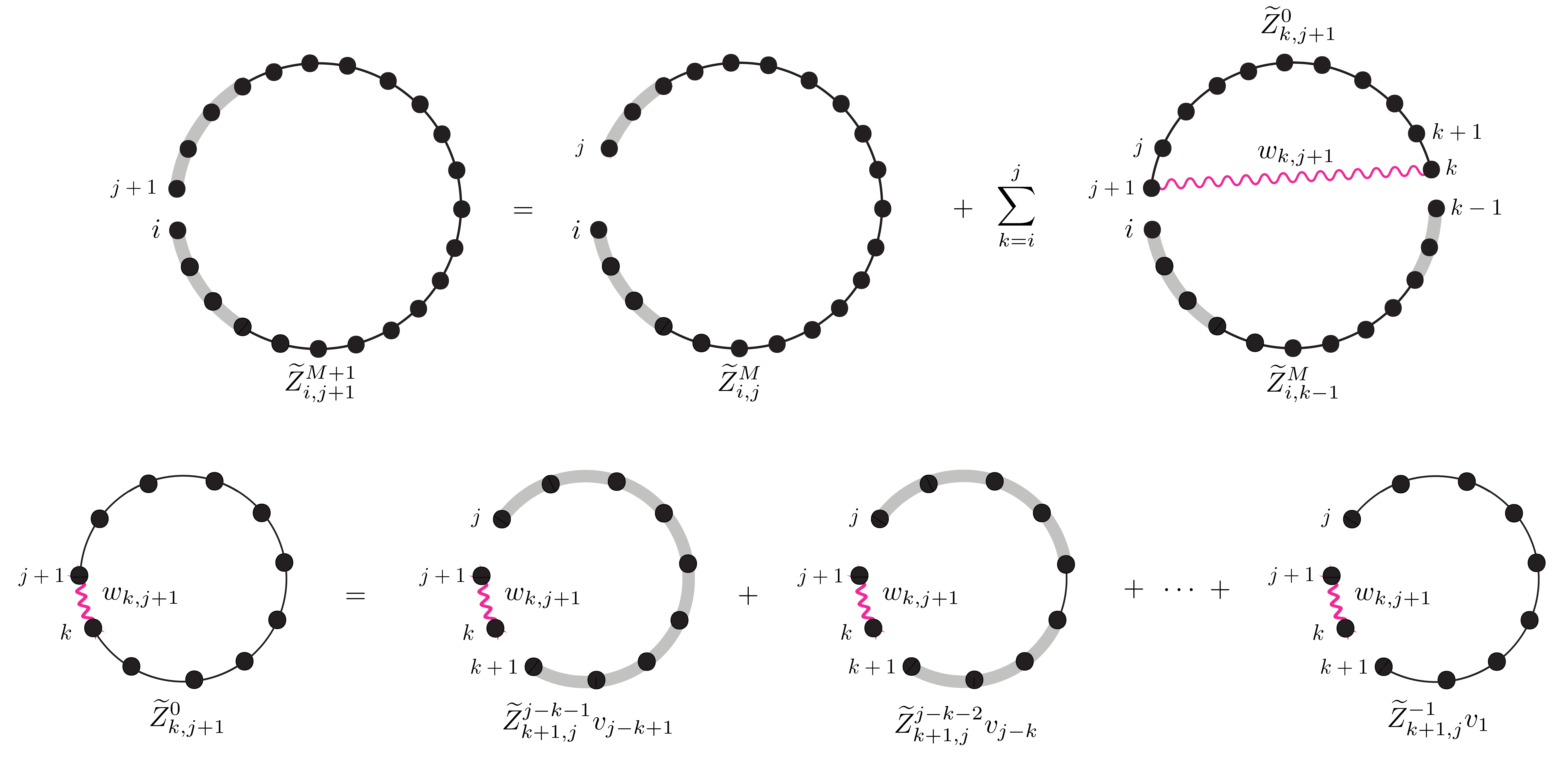}
  \caption{(Color online) Recursion scheme for the canonical partition function eq. (\ref{secext}). The partition function of a strand ranging from $i$ to $j+1$ with $M+1$  non-nested backbones  (thick grey lines) is computed from the strand ranging from $i$ to $j$ with $M$ non-nested backbones by adding a nested base at position $j+1$ and considering all possible pairings with a base $k\in (i,j)$. Each of these parings defines a structure which has zero non-nested backbones, i.e. an arbitrary substrand that is terminated by a helix. The explicit diagrams for the latter are showed in the second row and are obtained by considering  the sum over all non-nested backbones with associated statistical weight given by eq. (\ref{vu}).}
  \label{recursion}
\end{figure*} 

Einert \emph{et al.}  have shown how to take into account loops in the recursive equation by including the statistical weight 
\begin{eqnarray}
v_n=n^{-c},
\label{vu}
\end{eqnarray}
for each loop consisting of  $n$ links \cite{poland_scheraga} and summing over the restricted partition functions of strands terminated by a helix. 
If $Z_{i,j}^M$ denotes the partition function of a polymer going from monomer $i$ to monomer $j$ with  $M \le j-i$ unlooped links, see Fig. \ref{backbone}, given the initial conditions $Z_{i,i}^M = \delta_M$, $Z_{i,i-1}^M = \delta_{M+1}$, and $Z_{i,j}^{-1} = \delta_{j+1-i}$, the recursive partition function reads \cite{einert2}
\begin{eqnarray}
\widetilde{Z}_{i,j+1}^{M+1} 
=
 \widetilde{Z}_{i,j}^{M} + \sum_{k = i}^{j} w_{k,j+1} \widetilde{Z}_{i,k-1}^M \widetilde{Z}_{k,j+1}^0,
\label{secext}
\end{eqnarray} 
as illustrated in Fig. \ref{recursion}, with computational time $\mathcal{O}(N^4)$.
Here $\widetilde{Z}_{i,j}^M \equiv {Z_{i,j}^M}/{u_M}$ is the partition function rescaled by the statistical weight of $M$ non-looped links $u_M$, which will be not considered from this point onward, and  
\begin{eqnarray}
\widetilde{Z}_{k,j+1}^0 = \sum_{l = -1}^{j-k-1}   \widetilde{Z}_{k+1,j}^l  v_{l+2}
\end{eqnarray}
is the partition function of the arbitrary substrand that is terminated by a helix.

\subsection{Helicity degree}

\begin{figure}[h!]
\centering
\subfigure[]{\includegraphics[scale=0.35]{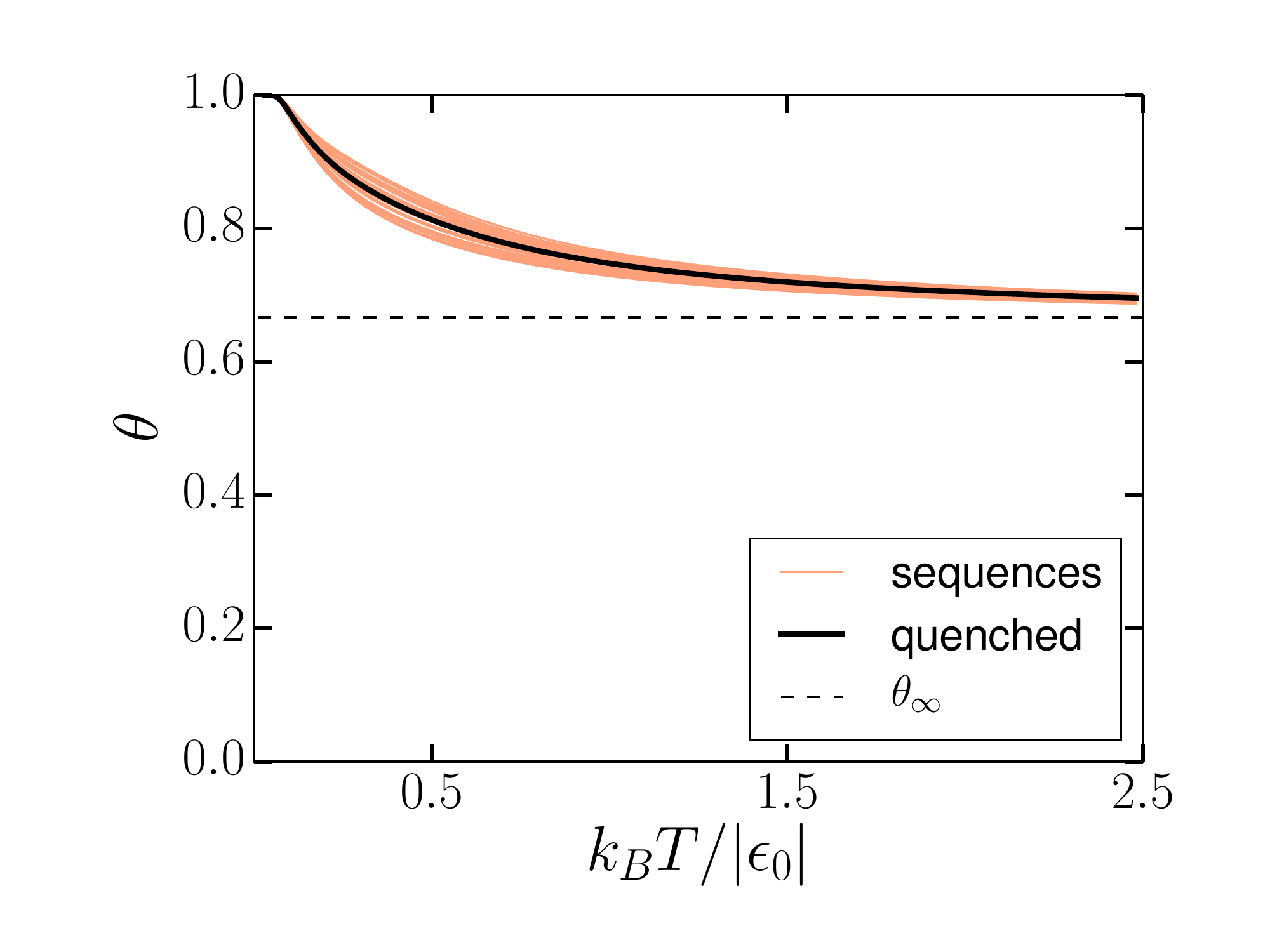}}
\subfigure[]{\includegraphics[scale=0.35]{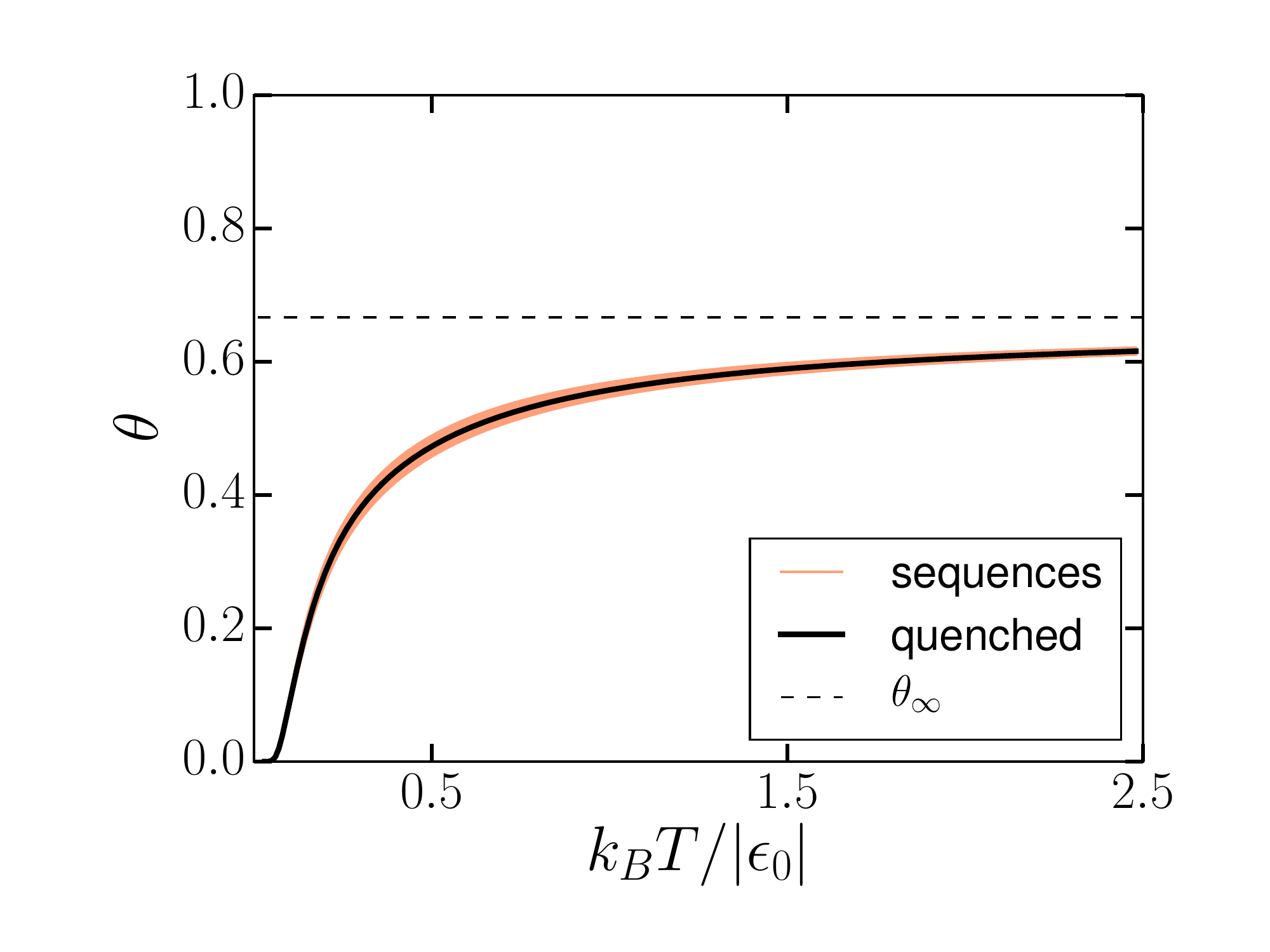}}
\caption{\label{quenched_eps0=+-1_eps=+-05} (Color online) Helicity degree for (a)  $\epsilon_0=-1$, $\epsilon=0.5|\epsilon_0|$ and for (b) $\epsilon_0=+1$, $\epsilon=-0.5|\epsilon_0|$ in the no-loop scenario with probability of U base occurrence $p=0.75$. Each red line corresponds to the helicity of a single RNA sequence realization of the disorder. The quenched average (black line) is obtained from the exact computation of the partition function for $30$ random sequences of length $N=50$. The black dashed line is the asymptote $\theta_{\infty}$ for $T\rightarrow \infty$ of eq. (\ref{23}).}
\end{figure}

The most relevant quantity that characterizes the conformation of the molecule is  the helicity degree defined as
\begin{eqnarray}
\theta = \frac{2}{N} \sum_{i<j} \ba{\braket{s_{i,j}}}= \frac{2}{N} \ba{\braket{\sum_{i<j} s_{i,j}}} = \frac{2}{N} \ba{\braket{|\mathcal{S}|}},
\label{theta}
\end{eqnarray}
where $\braket{...}$ denotes the average over the canonical ensemble and 
\begin{eqnarray}
|\mathcal{S}| \equiv \sum_{i<j} s_{i,j}
\end{eqnarray}
is the number of paired bases in the structure $\mathcal{S}$.
Since $\ba{\braket{|\mathcal{S}|}} \in [0,{N}/{2}]$, the helicity degree is a function of the temperature in the interval [0,1] with $\theta =1$ if every base is paired, corresponding to the native state, and $\theta =0$ if no base is paired. The helicity is a good measure of the order of the molecule conformation, since it gives the statistical weight of paired bases and thus can be used to quantify to what extent the molecule is folded. This can also be expressed in terms of the free energy by noting that
\begin{eqnarray}
\theta & & = 
\frac{2}{N} \ba{ \frac{1}{Z_N ( h )} \sum_{\{\mathcal{S}\}} e^{-\beta \epsilon_0 |\mathcal{S}|} e^{-\beta \epsilon \sum_{i < j} s_{i,j}h_i h_j}  |\mathcal{S}|}
\nonumber \\
& & =
\frac{2}{N} \ba{ \left(-\frac{\de}{\de(\beta\epsilon_0)}\right) \ln \sum_{\{\mathcal{S}\}} e^{-\beta \epsilon_0 |\mathcal{S}|} e^{-\beta \epsilon \sum_{i < j} s_{i,j}h_i h_j} }
\nonumber \\
& & =
2\frac{\de}{\de(\beta\epsilon_0)}\beta \ba{f(h)}.
\label{lunga}
\end{eqnarray}
Numerically the helicity can be estimated  in the absence of a loop free energy  using the probability of base pair formation $P_{i,j}$ between nucleotides $i$ and $j$  \cite{bund}
\begin{eqnarray}
P_{i,j} = \braket{s_{i,j}},
\qquad \Rightarrow \qquad 
\theta = \frac{2}{N} \ba{\sum_{i<j} P_{i,j}},
\end{eqnarray}
where the binding probability is obtained from the recursive equation as \cite{monthus}
\begin{eqnarray}
P_{i,j} =  e^{-\beta \epsilon_{i,j}} \frac{ Z_{i,j}^{int}Z_{i,j}^{ext}}{Z_{1,N}}.
\label{probbindembrio}
\end{eqnarray}
Here $Z_{i,j}^{int}$ is given by the partition function of the internal sequence $(i+1,\dots, j-1)$, while $Z_{i,j}^{ext}$ is the partition function of the external sequence 
$(1,\dots, i-1, j+1,\dots N)$. The latter can be computed by extending the recursion relation to the duplicated sequence $(1,\dots,N, N+1, \dots 2N)$ as $Z_{j+1,N+i-1}$ so that\begin{eqnarray}
P_{i,j} =  e^{-\beta \epsilon_{i,j}} \frac{ Z_{i+1,j-1}Z_{j+1,N+i-1}}{Z_{1,N}}.
\label{probbind}
\end{eqnarray}
The exact enumeration of $30$ random sequences, through the partition function with no loops eq. (\ref{recupart}), and the quenched average of the helicity degree  are shown in Fig. \ref{quenched_eps0=+-1_eps=+-05}, where we compare the attractive and repulsive background energy scenarios for the disordered two state model.  

\section{Homopolymer results}

Most of the known results for RNA belong to the special case of homopolymer models which we shall review in this section.
De Gennes was the first to obtain an expression for the the canonical partition function starting from the singularity analysis of the generating function using a propagator formalism \cite{degennes2}.
By setting $\epsilon = 0$ in eq. (\ref{hamiltonian}) the energy associated to each pairing becomes site independent, i.e.   $\epsilon_{i,j} = \epsilon_{0}$,  $\forall (i,j)$, thus making the energy of the structure $\mathcal{S}$ depending only on the number of paired bases $|\mathcal{S}|$.
  
\subsection{Folded RNA without loops}
Since  the restricted partition function is translationally invariant, in the no-loop scenario we can write $Z_{i,j} =  Q_{j-i+1}$ as the partition function of a homopolymer of length ${j-i+1}$ which depends only on the difference between position $i$ and position $j$. 
To decouple the summation it is useful to introduce the z-transform of $Q_N$
\begin{eqnarray}
Q(z) = \sum_{N=1}^\infty z^{-N}Q_N .
\end{eqnarray}
The canonical partition function $Z_N=Q_{N}$ is then obtained by back transforming the positive root of the resulting equation for  $Q(z)$ and in the limit of large $N$ a saddle point approximation yields \cite{bund}
\begin{eqnarray}
Z_N  = \sum_{\{\mathcal{S}\}} w^{|\mathcal{S}|}  \sim 
\xi(w)N^{\alpha-1} z_0^{-N},
\label{33}
\end{eqnarray}
where $\alpha = -1/2$ and $z_0 = 1/(1+2\sqrt{w})$, while $\xi(w)$ is a scaling function of the homopolymeric weight 
\begin{eqnarray}
w=\exp\left(-\beta \epsilon_0\right).
\label{weightomo}
\end{eqnarray}
In this scenario the free energy assumes the same scaling for all temperatures with universal pre factor $3/2$, originally obtained by de Gennes \cite{degennes2}, characteristic of the folded state for polymers and therefore no phase transition takes place.
From eq. (\ref{33})  it becomes possible to express the helicity degree as a function of the statistical weight in a very intuitive form. Indeed by writing  the partition function as $Z_N  = \sum_{\{\mathcal{S}\}} e^{|\mathcal{S}|\ln w}$, we have
\begin{eqnarray}
\theta^{hom}
=
\frac{2}{N} \braket{|\mathcal{S}|} 
=
\frac{2}{N} \frac{1}{Z_N} \frac{\de}{\de \ln w} Z_N
=
\frac{2}{N} \frac{\de \ln Z_N}{\de \ln w}.
\label{theta2}
\end{eqnarray}
Then using $w \de /\de w =  \de / \de \ln w$, for $N\gg 1$,  the helicity in the folded state takes the simple form
\begin{eqnarray}
\theta^{hom} 
& & =
\frac{2}{N} w \frac{\de \ln Z_N}{\de w}
\nonumber \\
& & =
\frac{2}{N} w \frac{\de }{\de w} \left(\ln \xi(w) -\frac{3}{2} \ln N + N\ln(1+2\sqrt{w})\right)
\nonumber \\
& & \approx
\frac{2\sqrt{w}}{1+2\sqrt{w}},
\label{heldegennes}
\end{eqnarray}
which asymptotically approaches a constant value 
\begin{eqnarray}
\theta_\infty = \lim_{T\rightarrow \infty} \theta^{hom} 
=
\lim_{\beta \rightarrow 0}
\frac{2e^{-\beta \epsilon_0/2}}{1+2e^{-\beta \epsilon_0/2}}
=
\frac{2}{3}.
\label{23}
\end{eqnarray}

\subsection{RNA folding with loops}

As in the no-loops scenario for homopolymers we set $w_{k,j}=w$ so that eq. (\ref{secext}) reads
\begin{eqnarray}
\widetilde{Q}_{N+1}^{M+1} 
=
 \widetilde{Q}_{N}^{M} + w \sum_{k = 0}^{N} \sum_{l = -1}^{N-k-1} \frac{\widetilde{Q}_{k-1}^M  \widetilde{Q}_{N-k-1}^l}{(l+2)^{c}},
 \label{partitionQ}
\end{eqnarray}
where the rescaled homopolymeric partition function $\widetilde{Q}_N^M$ describes a polymer with $N$ links and $M$ non-looped links with $-1\le M \le N$.
In absence of external forces the canonical partition function, which includes loop structures, is obtained by summing over all backbones as $Z_N^{loop} = \sum_{M=0}^\infty \widetilde{Q}_N^M$ and  the grand-canonical partition function follows as
\begin{eqnarray}
\mathcal{Z}^{loop}(z) = \sum_{N=0}^\infty z^N Z_N^{loop} =  \sum_{N=0}^\infty \sum_{M=0}^\infty z^N  \widetilde{Q}_N^M,
\label{firstgrandc} 
\end{eqnarray}
where $z$ is the fugacity.
By performing the double sum $\sum_{N=0}^\infty z^N\sum_{M=0}^\infty$ on both sides of eq. (\ref{partitionQ}), after rearranging indices  one obtains \cite{einert1} 
\begin{eqnarray}
\mathcal{Z}^{loop}(z)  = \frac{\kappa(w,z)}{1-z\kappa(w,z)}, 
\label{grandcano} 
\end{eqnarray}
where  
\begin{eqnarray}
\kappa(w,z) \equiv 1+w\sum_{l=-1}^\infty \sum_{N = l}^{\infty}   \frac{ z^{N+2} \widetilde{Q}_{N}^l}{(l+2)^{c}}
\label{kappaformal}
\end{eqnarray}
is the grand-canonical partition function of RNA structures with zero non-nested backbone links, i.e. structures which consist of just one nucleotide or structures where the terminal bases are paired.
For $|z\kappa|<1$ the grand-canonical partition function  eq. (\ref{grandcano}) can be expanded into a geometric series as $\mathcal{Z}^{loop} = \sum_{M=0}^{\infty} z^M\kappa^{M+1}$, and  comparing the coefficients of the power series with eq. (\ref{firstgrandc}) leads to $\sum_{N=M}^\infty z^N \widetilde{Q}_N^M = z^M\kappa^{M+1}$. Using this relation, eq. (\ref{kappaformal}) can be written as
\begin{eqnarray}
\kappa (w,z) - 1
=
\lambda(z,\kappa(w,z)),
\label{12bis}
\end{eqnarray}
where 
\begin{eqnarray}
\lambda(z,\kappa(w,z))\equiv \frac{w}{\kappa}\text{Li}(c,z\kappa(w,z)),
\end{eqnarray}
and $\text{Li}(c,x) \equiv \sum_{n=1}^{\infty}  {x^n}{n^{-c}}$ is the polylogarithm \cite{specfun}. This relation yields the first constitutive equation of the homopolymers theory with loop entropy. 
Going back to the canonical ensemble from eq.~\eqref{grandcano} the partition function takes the  general form of eq.~\eqref{33} but with $\alpha$ and $z_0$ not determined univocally.
In fact, contrary to the no-loop scenario, now the grand-canonical partition function features two relevant singularities. These are the single pole $z_0 = z_p$, where the denominator of eq. (\ref{grandcano}) vanishes, and  the branch point $z_0 = z_b$ of the function $\kappa(w,z)$, characteristic of the unfolded and folded phase respectively.

\begin{figure}[]
\centering
\subfigure[]{\includegraphics[scale=0.35]{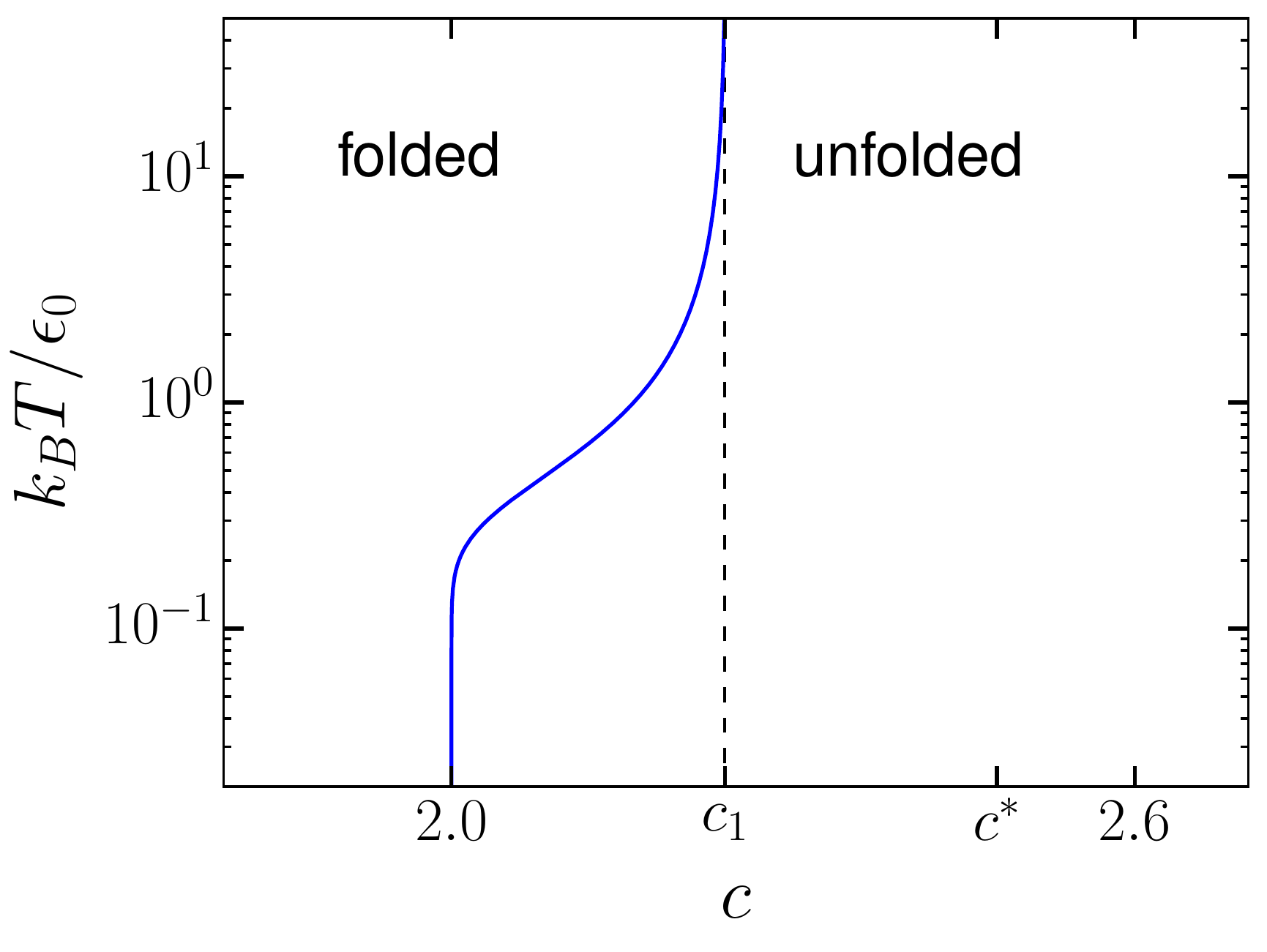}}
\subfigure[]{\includegraphics[scale=0.35]{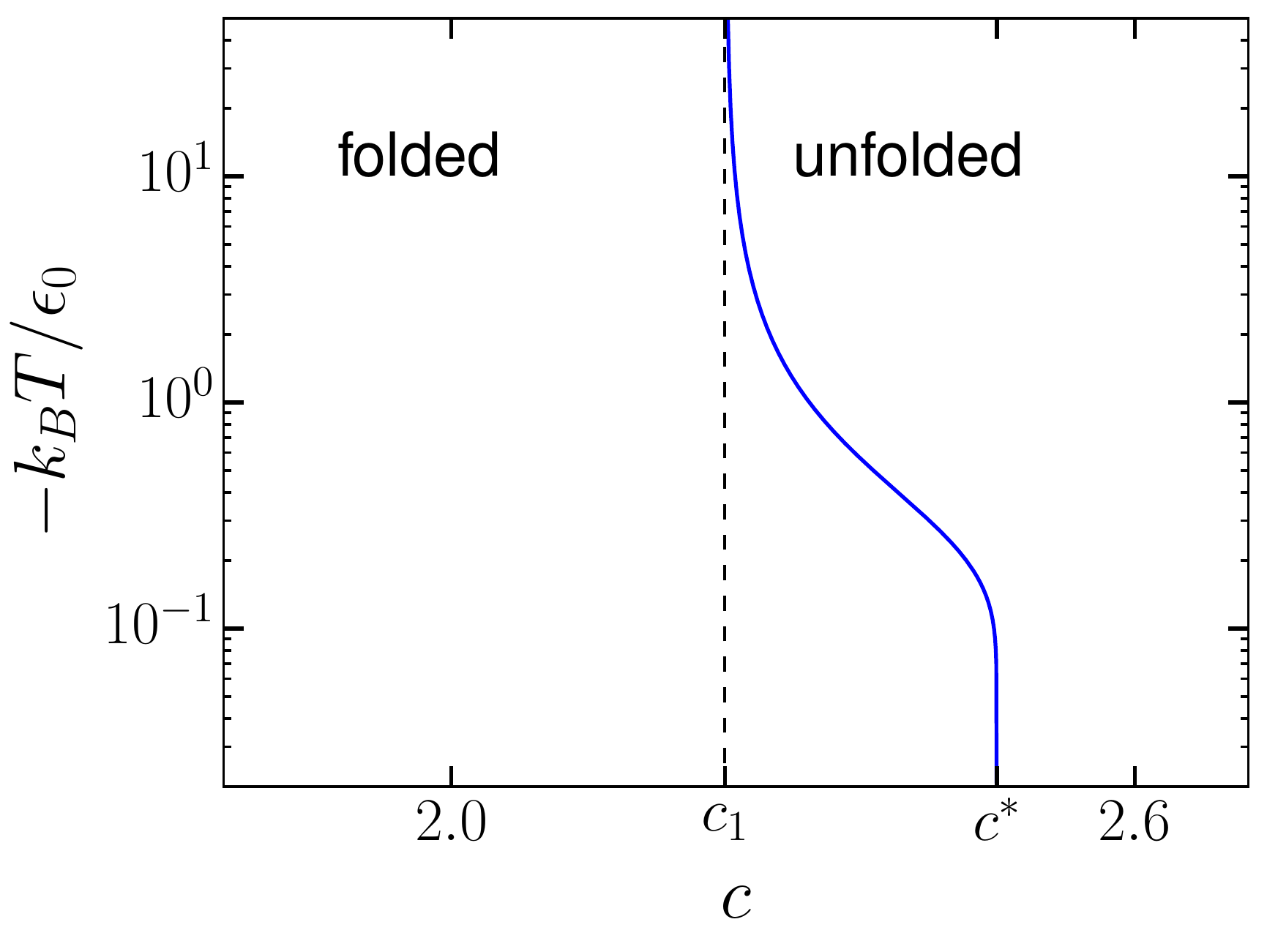}}
\caption{\label{wc}(Color online)  Phase diagram of homopolymeric RNA in the $T-c$ plane featuring the unfolded  and  folded phase for repulsive (a) and attractive (b) base pair interaction energy. The critical lines are obtained by solving $w_m (c)=w$ and in both cases they diverge for $c=c_1$. For $c\le 2$ the molecule is always folded and at $c=c^* \approx 2.479$ the critical weight $w_m (c)$ diverges so that for $c>c^*$ no folded phase can exist for both attractive and repulsive scenarios.}
\end{figure}

For $z<z_b$ at least one real solution of eq. (\ref{12bis}) exists, while exactly at $z=z_b$ the two solutions eventually merge and the slope of $\lambda(z,\kappa(w,z))$ at the tangent point $z_b$ equals unity. Imposing this condition, eq.  (\ref{12bis}) yields
\begin{eqnarray}
\kappa^2_b = w \left[\text{Li}(c-1,z_b\kappa_b) -  \text{Li}(c,z_b\kappa_b)\right],
\label{kbsq}
\end{eqnarray}
where we use the short notation $\kappa_b \equiv \kappa(w,z_b)$. This relation together with eq. (\ref{12bis}) univocally determines the branch singularity $z_b$, if it exists.
By a first order expansion of $\kappa(w,z)$ near the branch point the canonical partition function  scales as  
\begin{eqnarray}
Z_N^{loop} \sim  \xi_b(w) N^{-3/2} {z_b}^{-N},
\label{canoscalekb}
\end{eqnarray} 
which leads for the free energy to the logarithmic $N$-contribution with universal prefactor $3/2$, in agreement  with the no-loop scenario eq. (\ref{33}). This is therefore the partition function describing homopolymeric RNA with loop structures in the folded phase. From eq. (\ref{canoscalekb}) the helicity degree eq. (\ref{theta2}) follows as \cite{einert1}
\begin{eqnarray}
\theta_b
\approx 
\frac{2w}{N}  \frac{\de \ln z_b^{-N}}{\de w} 
=
-\frac{2w}{z_b} \frac{\de z_b }{\de w}
=
\frac{2 \text{Li}(c,z_b\kappa_b)}{\text{Li}(c-1,z_b\kappa_b)}.
\label{thetab}
\end{eqnarray} 
Instead the simple pole $z_p$ is determined, together with eq. (\ref{12bis}), by
\begin{eqnarray}
z_p \kappa_p = 1,
\label{szk}
\end{eqnarray} 
where $\kappa_p \equiv \kappa(w,z_p)$. By inserting this into eq. (\ref{12bis}) yields an explicit expression for the pole singularity as  
\begin{eqnarray}
z_p
=2 \left(1+\sqrt{1+4w\zeta(c) }\right)^{-1},
\label{zp}
\end{eqnarray} 
where we use the Riemann zeta function $\zeta(c) = \text{Li}(c,1) = \sum_{n=1}^\infty n^{-c}$. In this scenario the partition function scales as 
\begin{eqnarray}
Z_N^{loop} \sim  \xi_p(w) {z_p}^{-N},
\label{canoscalekp}
\end{eqnarray} 
which in contrast to the branch point does not lead to the logarithmic $N$-contribution for the free energy, since the singularity exponent is $\alpha=1$, and describes the thermodynamics of homopolymeric RNA above the melting critical  point. In this phase the helicity degree takes the form 
\begin{eqnarray}
\theta_p
\approx 
-\frac{2w}{z_p} \frac{\de z_p }{\de w}
=
1-\frac{1}{\sqrt{1+4w\zeta(c)}} .
\label{thetap}
\end{eqnarray} 
Since the critical behavior of the system is characterized by the singularity that is closest to the origin in the complex $z$ plane, a phase transition is possible only if a critical fugacity $z_{m}$ and a  critical weight $w_{m}$ exist such that $z_{m} = z_b(w_{m}) = z_p(w_{m})$.
Then, at the critical point the three constitutive equations (\ref{12bis}), (\ref{kbsq}) and (\ref{szk}) have to hold simultaneously. Using equations (\ref{szk})  and (\ref{zp}) a closed form expression can be given for the critical weight as a function of the loop exponent only 
\begin{eqnarray}
 w_m (c)= \frac{ \zeta(c-1) - \zeta(c) }{\left(\zeta(c-1) - 2\zeta(c)\right)^2}.
\label{wc2}
\end{eqnarray} 
This expression, which  determines the critical line in the homopolymeric phase diagram, defines an upper bound  for the range of the loop exponent $c$ in which a phase transition can occur. This is given by the universal value $c^*\approx 2.479$ at which the denominator of eq. (\ref{wc2}) vanishes and $w_m$ diverges so that only the pole singularity can exist and the molecule is always unfolded. 
Furthermore from the definition of polylogarithm it follows that the derivative with respect to $\kappa$ of the function $\lambda(z,\kappa)$ converges only for $c>2$, and since at $z=z_b$ this has to be equal to unity, this sets also a lower bound to the critical region in $c$.
Depending on the nature of the interaction, repulsive if $\epsilon_0 >0$ and attractive otherwise, the behavior of the statistical weight of base pair formation $w=\exp(-\beta \epsilon_0)$ can be smaller or greater than unity. Then one can define the  additional universal value of the loop exponent 
\begin{eqnarray}
c_1\approx 2.241,
\label{c1}
\end{eqnarray}
defined by 
\begin{eqnarray}
\zeta(c_1-1) - \zeta(c_1)  = \left(\zeta(c_1-1) - 2\zeta(c_1)\right)^2.
\label{wc2_c1}
\end{eqnarray} 
From this argument one concludes that only for $2 < c < c^*$ a phase transition can occur between the folded and the unfolded phase, determined by $z_b$ and $z_p$ respectively, since only then both singularities can coexist. The corresponding phase diagram for attractive and repulsive interaction is obtained by solving $w_m (c)=w$, see Fig. \ref{wc}.

\section{Constrained annealing with loop entropy}

\subsection{Outline of the method}

Instead of the standard replica approach used for spin glasses \cite{par}, to compute the average over the disorder we use the constrained annealing approximation \cite{serva}. The basic idea is to perform an annealed average, where the annealed free energy is defined as
\begin{eqnarray}
f^a =  -\frac{1}{\beta  N} \ln \ba{Z_N(h)},
\label{fan}
\end{eqnarray} 
with the random variables $\{h\}$ coupled to appropriate constraints $\{\mu\}$, which are functions of  some disorder self-averaging variables \cite{par}. The values of the constraints, which assume the form of Lagrange multipliers, that for $N\gg 1$ maximize the thermodynamic potential 
\begin{eqnarray}
f^{ca}(\mu) = - \frac{1}{\beta N}\ln Z_N^{ca}(\mu),
\end{eqnarray}
are those that select the realizations with a correct value of the disorder intensive variables and at the same time that minimize the difference between the quenched free energy eq. (\ref{fque}) and the annealed free energy eq. (\ref{fan}), in which the disorder variables $\{h\}$ are free to evolve as the dynamical degrees of freedom $\{s\}$. 
Thus $f^{ca}(\mu)$ improves the lower bound estimation of the Jensen's inequality \cite{monta} given by $f^a$ for the quenched free energy so that 
\begin{eqnarray}
\ba{f(h)} \ge f^{ca}(\mu) \ge f^a \qquad \forall \mu.
\end{eqnarray}
To construct the partition function \cite{serva}
\begin{eqnarray}
Z_N^{ca}(\mu) =  \ba{Z_N(h) e^{-N\mu\alpha (h)}},
\label{Zca}
\end{eqnarray}
we define a function of the disordered sequence which self-averages to zero as 
\begin{eqnarray}
\alpha(h) \equiv \frac{1}{N}  \sum_{i = 1}^N [h_i - (2p-1)].
\end{eqnarray}
It follows immediately from
\begin{eqnarray}
\ba{h} 
\equiv
\sum_{\{h\}}  \mathcal{P}(h) h 
=
\sum_{h=\pm1} \rho (h) h
=
2p-1,
\label{hbar}
\end{eqnarray}
that $\alpha(h)\rightarrow 0$ for $N\rightarrow \infty$. 
As we will show next, since the Hamiltonian  (\ref{hamiltonian}) is  separable \cite{bascle} as we consider site instead of link random variables,  with this choice of  $\alpha(h)$ in the thermodynamic limit the disorder terms an be averaged independently.
To see this we write eq. (\ref{partfun}) as
\begin{eqnarray}
Z_N ( h ) 
= 
\sum_{\{\mathcal{S}\}} e^{-\beta \epsilon_0 |\mathcal{S}|} \prod_{k<l}^Ne^{-\beta \epsilon s_{k,l}h_k h_l},
\end{eqnarray}
so that the constrained annealing partition function eq. (\ref{Zca}) becomes
\begin{eqnarray}
Z_N^{ca}(\mu) =
e^{N\mu(2p-1)}\sum_{\{\mathcal{S}\}} e^{-\beta \epsilon_0 |\mathcal{S}|} \Pi (\mu),
\label{zmezzo}
\end{eqnarray}
with
\begin{eqnarray}
 \Pi (\mu) \equiv  \sum_{\{h\}} \prod_{i=1}^N \rho (h_i)e^{-\mu h_i} \prod_{k<l}^Ne^{-\beta \epsilon s_{k,l}h_k h_l}.
\end{eqnarray}
At this point the key observation is that since in the product we have  contributions different from unity only when $s_{k,l} =1$ and each base can only participate in at most one base pair, in the average over the disorder we get a product of $|\mathcal{S}|=\sum_{i<j} s_{i,j}$ times the factor $e^{-\beta \epsilon h_k h_l}$. This is equivalent to saying that every disorder term  can be averaged independently, which  follows as a direct consequence of the mutual independence of the sequence disorder variables $h_i$. Thus explicating the summation we obtain 
\begin{eqnarray}
 \Pi (\mu)
& = &
\left(\sum_{h=\pm 1} \rho (h)e^{-\mu h}\right)^{N-2|\mathcal{S}|} \times
\nonumber \\
& & \times \left(\sum_{h,h'=\pm 1} \rho (h)\rho (h')e^{-\mu h}e^{-\mu h'}e^{-\beta \epsilon h h'}\right)^{|\mathcal{S}|}
\nonumber \\
&  = &
\Omega^{N}(\mu) \left(\frac{\Upsilon(\mu) }{\Omega^2(\mu) }\right)^{|\mathcal{S}|},
\end{eqnarray}
where we have defined the two auxiliary quantities $\Omega (\mu)\equiv p e^{-\mu} +(1-p)e^\mu$ and  $\Upsilon (\mu) \equiv e^{-\beta \epsilon}[p^2 e^{-2\mu} + (1-p)^2 e^{2\mu}] + e^{\beta \epsilon}2p(1-p)$.  This can be written in a more compact form if we define a new constant interaction energy 
\begin{eqnarray}
\ba{\epsilon}(\mu)  \equiv - \frac{1}{\beta}\ln \frac{\Upsilon(\mu) }{ \Omega^2(\mu) },
\label{epsbar}
\end{eqnarray}
where all the information relative to  the disorder average is now included in the parameter on which to perform the variation $\mu$.
Then eq. (\ref{zmezzo}) reduces to
\begin{eqnarray}
Z^{ca}_N (\mu)= e^{N\mu(2p-1)}\Omega^{N}(\mu) Z^{hom}_N(\mu),
\label{capartfun}
\end{eqnarray}
where $Z^{hom}_N (\mu) = \sum_{\{\mathcal{S}\}} [w^{ca}(\mu)]^{|\mathcal{S}|}$ is a homopolymeric partition function associated with the constrained annealing weight  
\begin{eqnarray}
w^{ca}(\mu) =  \exp\left[-\beta \epsilon^{ca}(\mu)\right],
\end{eqnarray}
with pair interaction energy
\begin{eqnarray}
\epsilon^{ca}(\mu) = \epsilon_0+\ba{\epsilon}(\mu).
\label{epsca}
\end{eqnarray}
From its definition in the limit of vanishing disorder strength $\epsilon \rightarrow 0$ one simply recover the homopolymeric statistical weight of eq. (\ref{weightomo}) with constant interaction energy $\epsilon^{ca} = \epsilon_0$.
The maximisation with respect to $\mu$  of 
\begin{eqnarray}
\beta f^{ca}(\mu) = -\mu(2p-1) - \ln \Omega(\mu) - \frac{1}{N}\ln Z^{hom}_N (\mu)
\end{eqnarray}
is achieved by imposing
\begin{eqnarray}
\frac{\de}{\de \mu} f^{ca}(\mu) \bigg|_{\mu = \widetilde{\mu}}= 0.
\label{dedemu}
\end{eqnarray}
This condition yields the value $\widetilde{\mu}(\beta\epsilon_0,\beta\epsilon)$ for which 
\begin{eqnarray}
f^{ca}(\widetilde{\mu}) = \max_{\mu} f^{ca}(\mu) \approx \ba{f(h)},
\end{eqnarray}
while the annealed free energy  is obtained by setting $\mu=0$ in $Z^{ca}_N(\mu)$.

\section{Critical behaviour}

The behaviour of the specific heat  and helicity degree has been addressed recently by Hayrapetyan \emph{et al.} \cite{roland1} by solving eq. (\ref{dedemu}), where eq. (\ref{33}) for $Z^{hom}_N$ is used, with results showing a very good agreement between the quenched  and constrained annealing averages.
By changing the disorder strength $\epsilon$, they showed that in the double peak structure of $C_V$ obtained for $0.5<p<1$ and $\epsilon_0\ne 0$, the low temperature jump is more pronounced than the high temperature one when $\theta$ decreases from its maximum value as $T\rightarrow 0$ and vice versa.
Instead, a partition function with $\epsilon_0=0$ yields a single peak structure for all values of the disorder strength $\epsilon$. 

In this paper we account for loop entropy by using the scalings equations (\ref{canoscalekb}) and (\ref{canoscalekp}) with corresponding free energies 
\begin{eqnarray}
\beta f^{ca}_{[b,p]}(\mu) \approx -\mu(2p-1) - \ln \Omega(\mu) +\ln z_{[b,p]}(\mu),
\label{fcab}
\end{eqnarray}
where 
\begin{eqnarray}
z_p(\mu)
=2 \left(1+\sqrt{1+4w^{ca} ({\mu})\zeta(c) }\right)^{-1}.
\label{zp2}
\end{eqnarray} 
Thus a thermal  phase transition is triggered by  the different nature of the two singularities in the homopolymeric partition function and is therefore expected to explain physically the unusual drop of $\theta$ at low temperatures found in \cite{roland1}. By keeping implicit the expression for $Z^{hom}_N$,  eq. (\ref{dedemu}) yields 
\begin{eqnarray}
0 & = &
2p-1 + \bigg[\frac{\de \ln \Omega(\mu)}{\de \mu}  + \frac{1}{N Z^{hom}_N(\mu)} \times
\nonumber \\
& & \times 
\sum_{\{\mathcal{S}\}} |\mathcal{S}| (w^{ca}(\mu))^{|\mathcal{S}|-1} \frac{\de }{\de \mu} w^{ca} (\mu)\bigg]_{\mu = \widetilde{\mu}}
\nonumber \\
& = & 
2p-1 + \bigg[\frac{\de \ln \Omega(\mu)}{\de \mu}  +  \frac{\braket{|\mathcal{S}|}}{N}  \frac{\de \ln w^{ca} (\mu)}{\de \mu} 
 \bigg]_{\mu = \widetilde{\mu}}
  \nonumber \\
 & = & 
2p-1 + \bigg[\frac{\de \ln \Omega(\mu)}{\de \mu}  +  \frac{\theta^{ca}(\mu) }{2}  \frac{\de }{\de \mu} \ln \frac{\Upsilon(\mu) }{ \Omega^2(\mu) }
 \bigg]_{\mu = \widetilde{\mu}}
 \label{troppo}
\end{eqnarray}
where we have  used equations (\ref{theta2}),  (\ref{epsbar}) and  (\ref{epsca}), and where
\begin{eqnarray}
\theta^{ca}(\mu) & & = 
2\frac{\de}{\de(\beta\epsilon_0)}\ln \left(\frac{1}{1+2\sqrt{w^{ca}(\mu)}}\right)
\nonumber \\
& & =
 \frac{-2}{1+2\sqrt{w^{ca}(\mu)}}\left(\frac{1}{\sqrt{w^{ca}(\mu)}}\frac{\de w^{ca}(\mu)}{\de(\beta\epsilon_0)}\right)
 \nonumber \\
& & =
\frac{2\sqrt{w^{ca}({\mu})}}{1+2\sqrt{w^{ca}({\mu})}}.
\label{tetacab}
\end{eqnarray}

\begin{figure}[]
	\centering
	\subfigure[]{\includegraphics[width=0.33\textwidth]{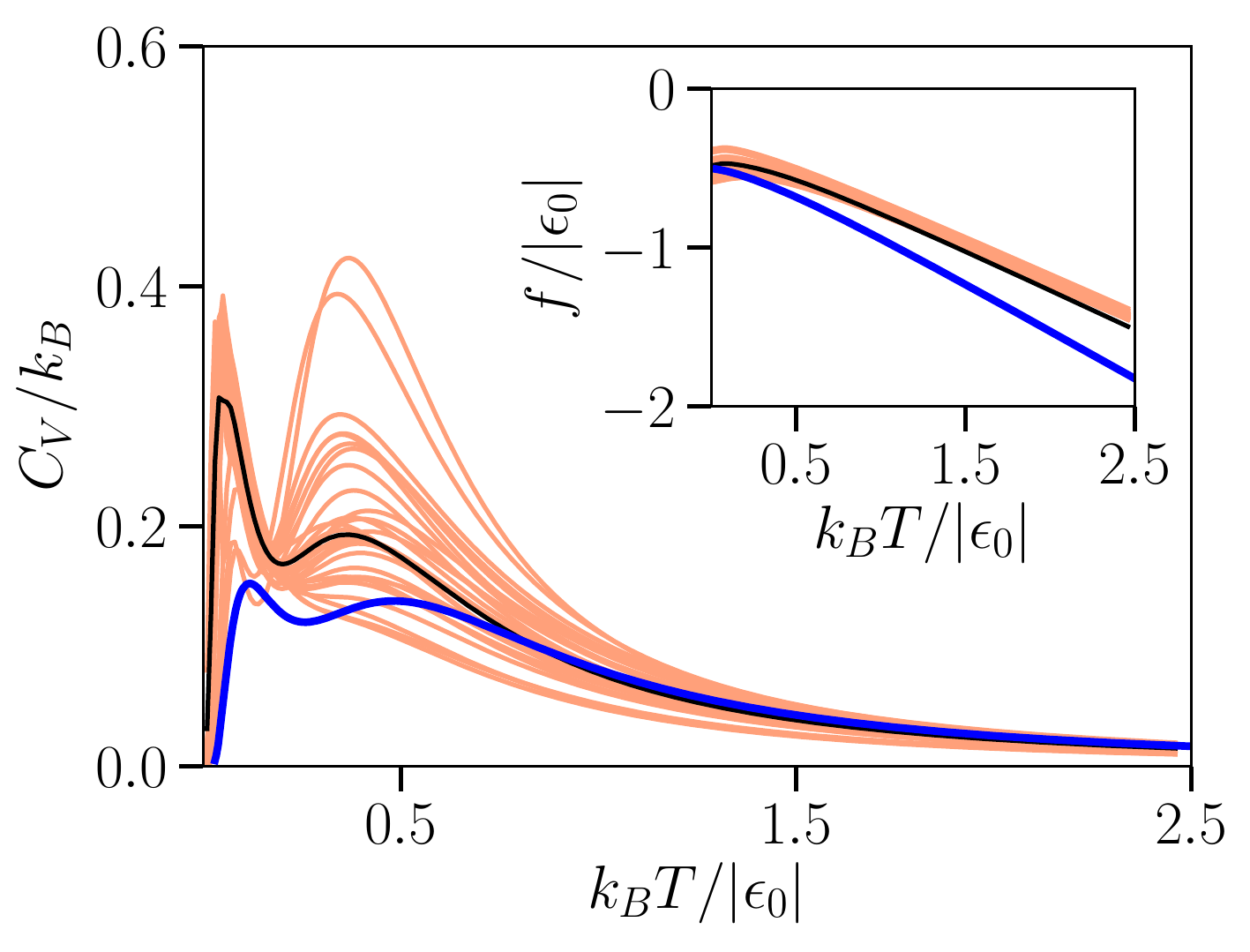}}
	\subfigure[]{\includegraphics[width=0.33\textwidth]{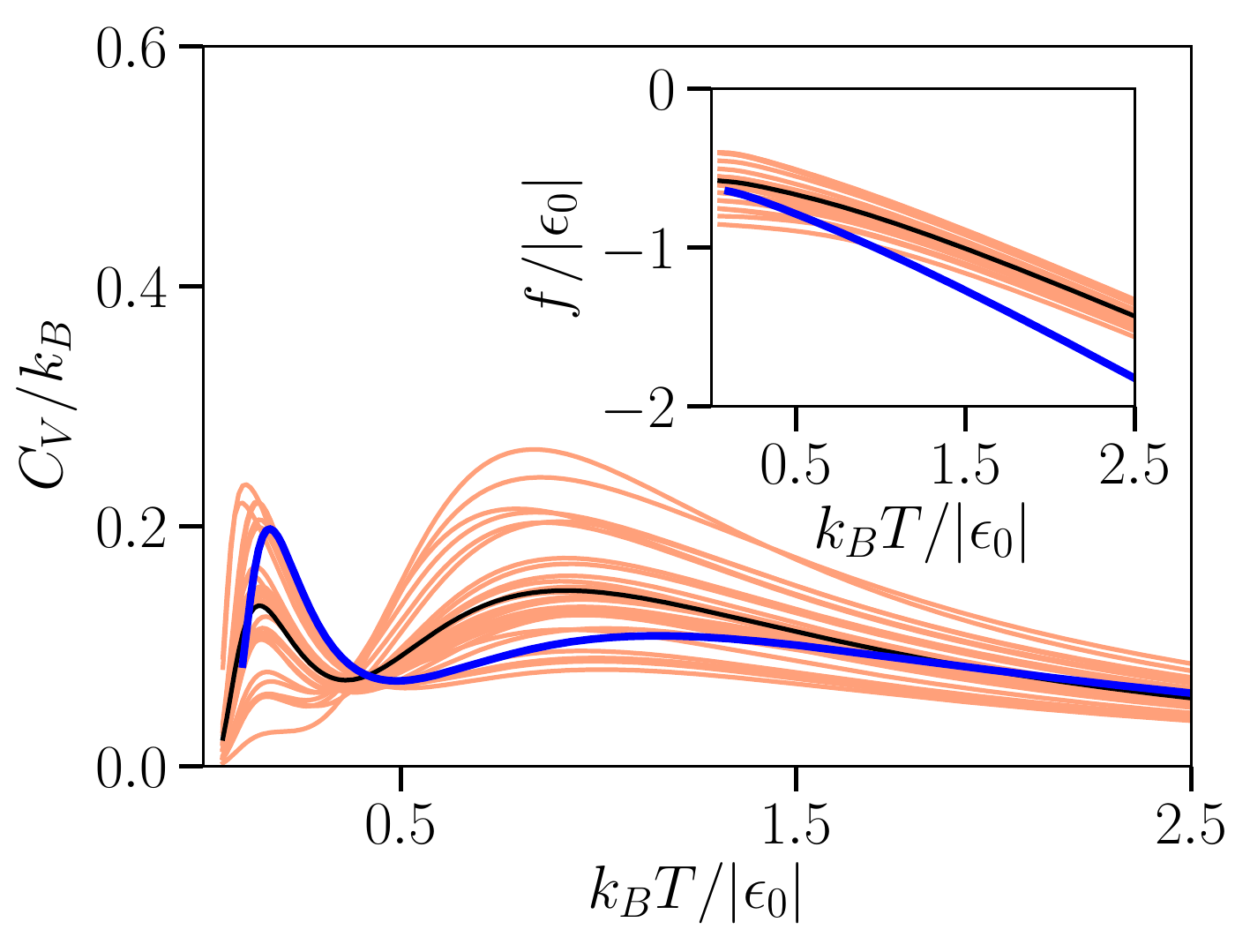}}
	\subfigure[]{\includegraphics[width=0.33\textwidth]{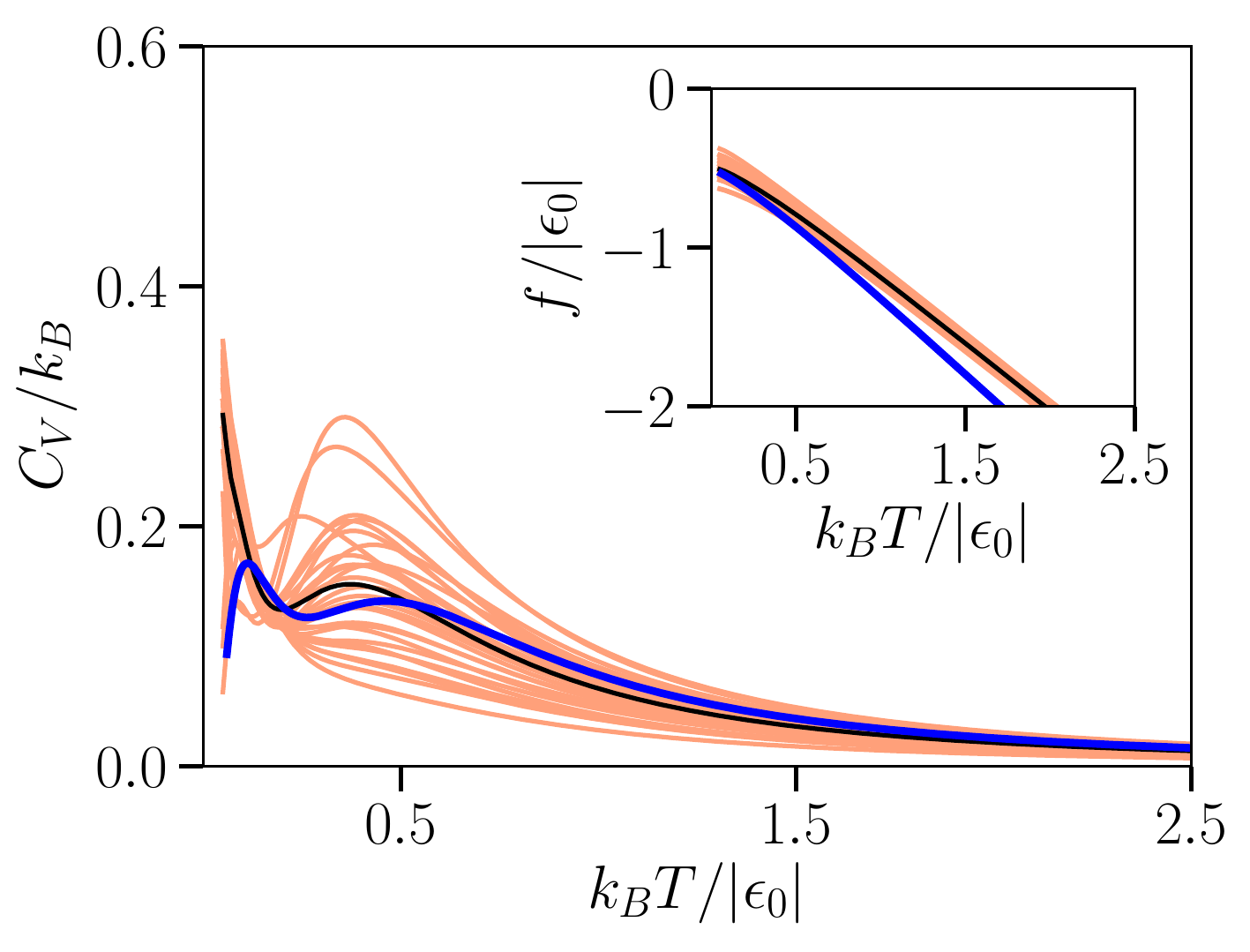}}
	\subfigure[]{\includegraphics[width=0.33\textwidth]{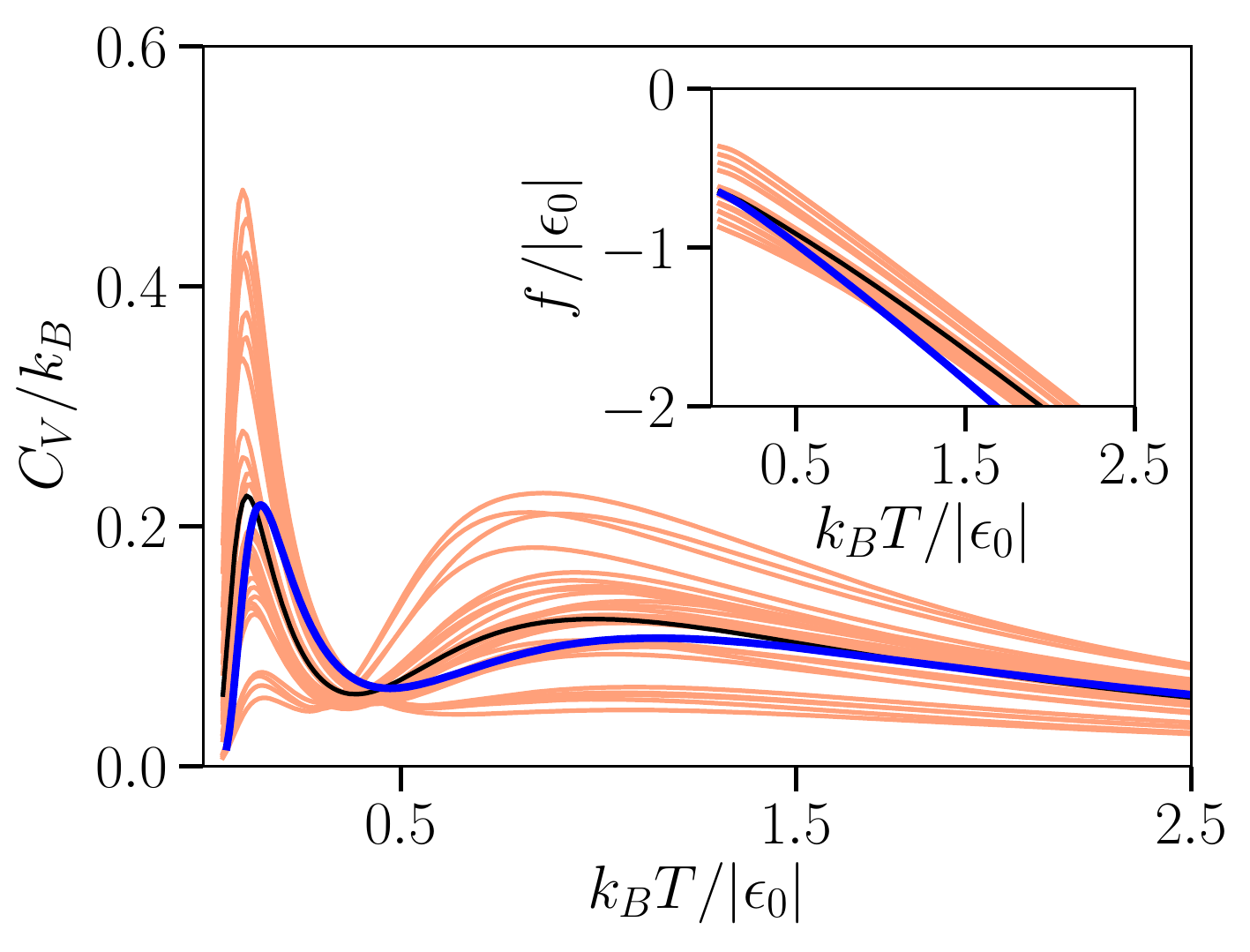}}
	\caption{\label{quenched} (Color online) Folded phase specific heat with $c=c_{RW} = 1.5$ (a,b) and without loop entropy $c=0$ (c,d). The quenched average (black line) is obtained from $30$ random sequences (red lines) with $N=50$, $p=0.75$,  (a) $\epsilon=0.5|\epsilon_0|$ and (b) $\epsilon=1.5|\epsilon_0|$. In  blue the  constrained annealing average.  Insets: the corresponding free energies with quenched and constrained annealing averages from which the specific heat have been obtained.}
\end{figure}

\begin{figure}[]
\centering
\includegraphics[scale=0.35]{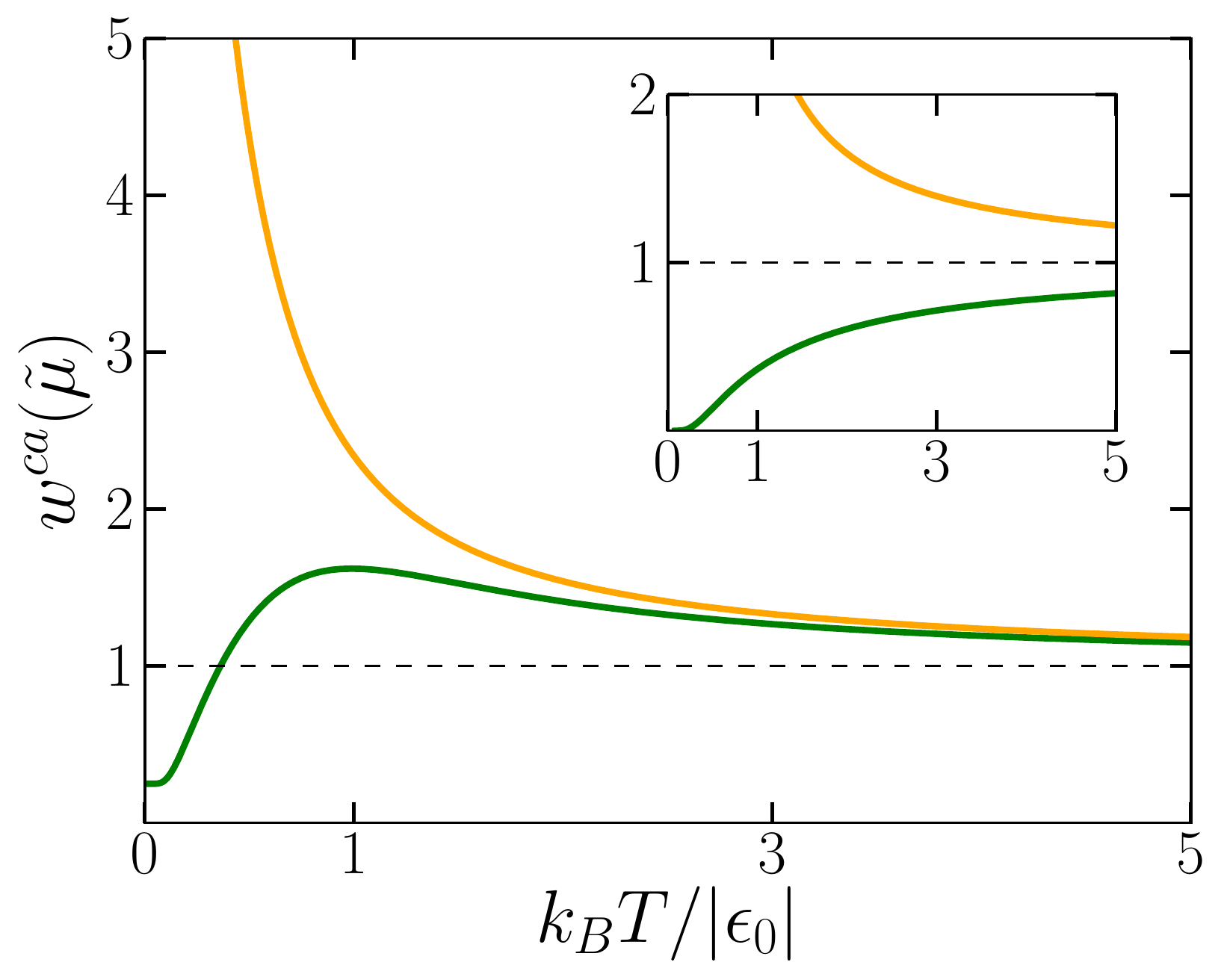}
\caption{\label{weight} (Color online) Constrained annealing weight in the competitive scenario ($\Lambda<1$) for $p=0.75$, $\epsilon = 1.5|\epsilon_0|$ (green line) and  in the non-competitive scenario ($\Lambda>1$) for $p=0.75$, $\epsilon = 0.5|\epsilon_0|$ (orange line)  as a function of the temperature. Inset: Homopolymeric weight $w=e^{-\beta \epsilon_0}$ with $ \epsilon_0>0$ (green line) and $ \epsilon_0<0$ (orange line), for repulsive and attractive regimes respectively.}
\end{figure}

\subsection{Cold melting} 

Once the solution of eq. (\ref{troppo}) is known, the folded phase singularity $z_b(\widetilde{\mu})$ can be obtained by solving the two equations 
\begin{eqnarray}
\begin{cases}
\widetilde{\kappa_b^{ca}} = \frac{w^{ca} (\widetilde{\mu})}{\widetilde{\kappa_b^{ca}}}\text{Li}(c,z_b(\widetilde{\mu})\widetilde{\kappa_b^{ca}})+1
\\
\\
\widetilde{\kappa_b^{ca}} = \frac{w^{ca}(\widetilde{\mu})}{\widetilde{\kappa_b^{ca}}} 
\left[
\text{Li}(c-1,z_b(\widetilde{\mu})\widetilde{\kappa_b^{ca}}) 
- \text{Li}(c,z_b(\widetilde{\mu})\widetilde{\kappa_b^{ca}})
\right]
\end{cases}
\label{kbsq2}
\end{eqnarray}
where $\widetilde{\kappa_b^{ca}}\equiv \kappa(w^{ca} (\widetilde{\mu}),z_b(\widetilde{\mu}))$.
In the unfolded phase $z_p(\widetilde{\mu})$ is instead determined by the explicit expression eq. (\ref{zp2}) for $\mu = \widetilde{\mu}(\beta \epsilon_0, \beta \epsilon)$. Then the free energies in the folded and unfolded phases are computed using eq. (\ref{fcab}).
The critical weight $w_m(c)$ defined by eq. (\ref{wc2}) is a  monotonically increasing function of the loop exponent $c$ in the interval $2 < c < c^*$ which defines the critical region for homopolymeric RNA. For $w>w_m$ the molecule is always folded, governed by $z_b$, and unfolded otherwise, governed by $z_p$. 
Then the critical temperature for the disordered model can be estimated numerically by imposing $w^{ca} (\widetilde{\mu}) = w_m(c)$ for fixed $c$, yielding the critical value of the Lagrange multipliers $\widetilde{\mu}_m(c)$. 

In Fig. \ref{quenched} we compare quenched and constrained annealing averages in the folded phase for  (a,b) $c=c_{RW} = 1.5$ which qualitatively reproduce the behavior for $c=0$ in (c,d), where a double peak structure was also found in the specific heat. 
Here, the partition function of each sequence is computed with the loop recursive equation (\ref{secext}) and the constrained annealing free energy is $f^{ca}_b(\widetilde{\mu})$ of eq. (\ref{fcab}) from which the specific heat  follows as
\begin{eqnarray}
C_V^{ca}(\widetilde{\mu})  = \frac{k_B T}{N} \frac{\de^2 T \ln Z_N^{ca}(\widetilde{\mu})}{\de T^2}= -T \frac{ \de^2  f^{ca}(\widetilde{\mu}) }{\de T^2}.
\end{eqnarray}
The quenched and constrained annealing free energies show a good agreement for low temperatures while increasing $T$ a gap arises.
This energy gap however seems to increase linearly in $T$  and does not affect the specific heat qualitatively.
For $\epsilon_0<0$, which account for an attractive background interaction, we also allow the helicity to reach its maximum value when $T\rightarrow 0$. For the following analysis it is useful to separate the two main energetic regimes depending on whether quenched disorder has an effective relevance in the global behavior.

\begin{figure}[]
\centering
\subfigure[]{\includegraphics[scale=0.45]{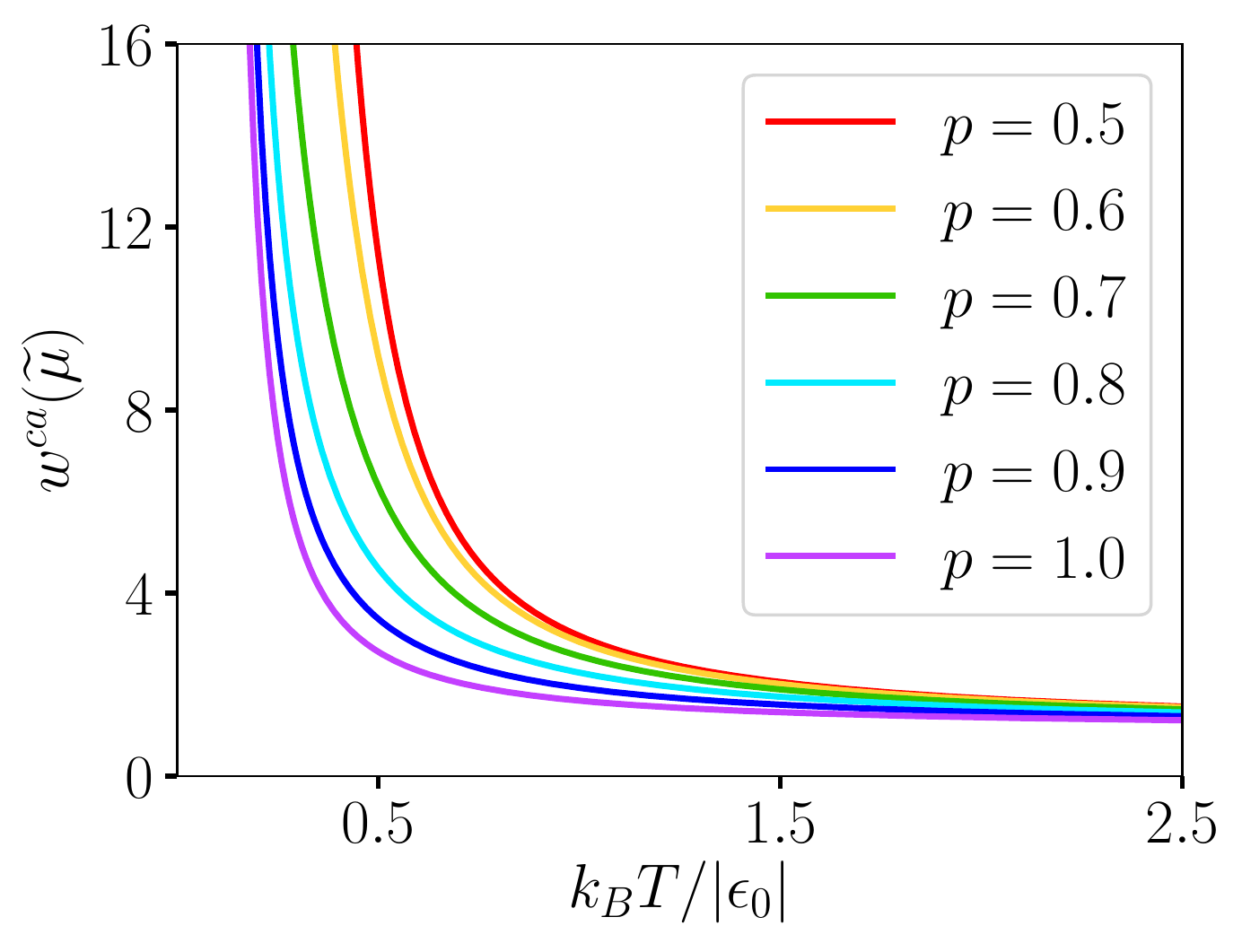}}
\subfigure[]{\includegraphics[scale=0.45]{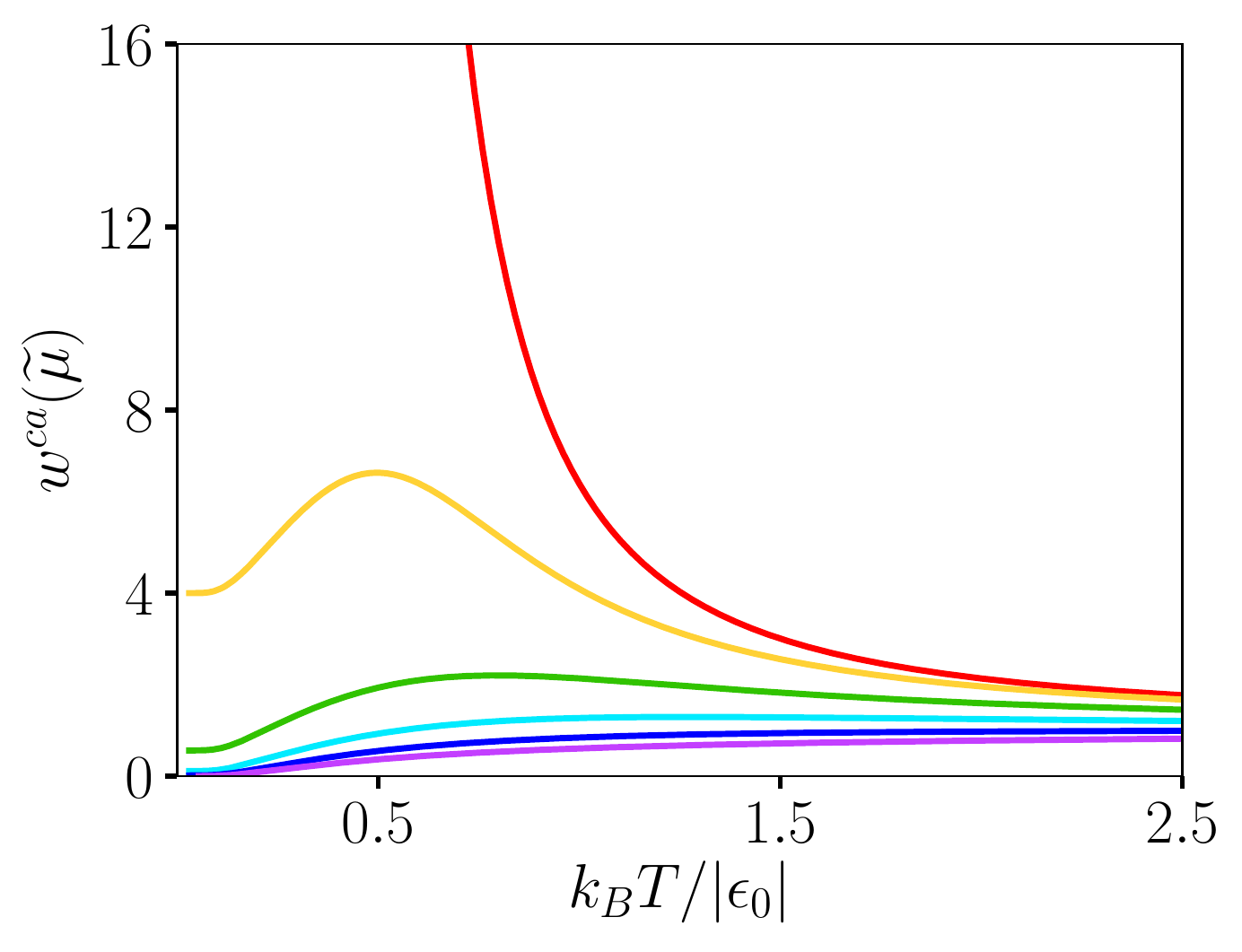}}
\caption{\label{consan} (Color online) Statistical weight at various probabilities $0.5\le p\le1.0$ in the constrained annealing approach (a) non-competitive scenario with $\Lambda = 2$ and  (b) competitive scenario with $\Lambda = 2/3$.}
\end{figure}

\begin{figure}[]
\centering
\subfigure[]{\includegraphics[scale=0.45]{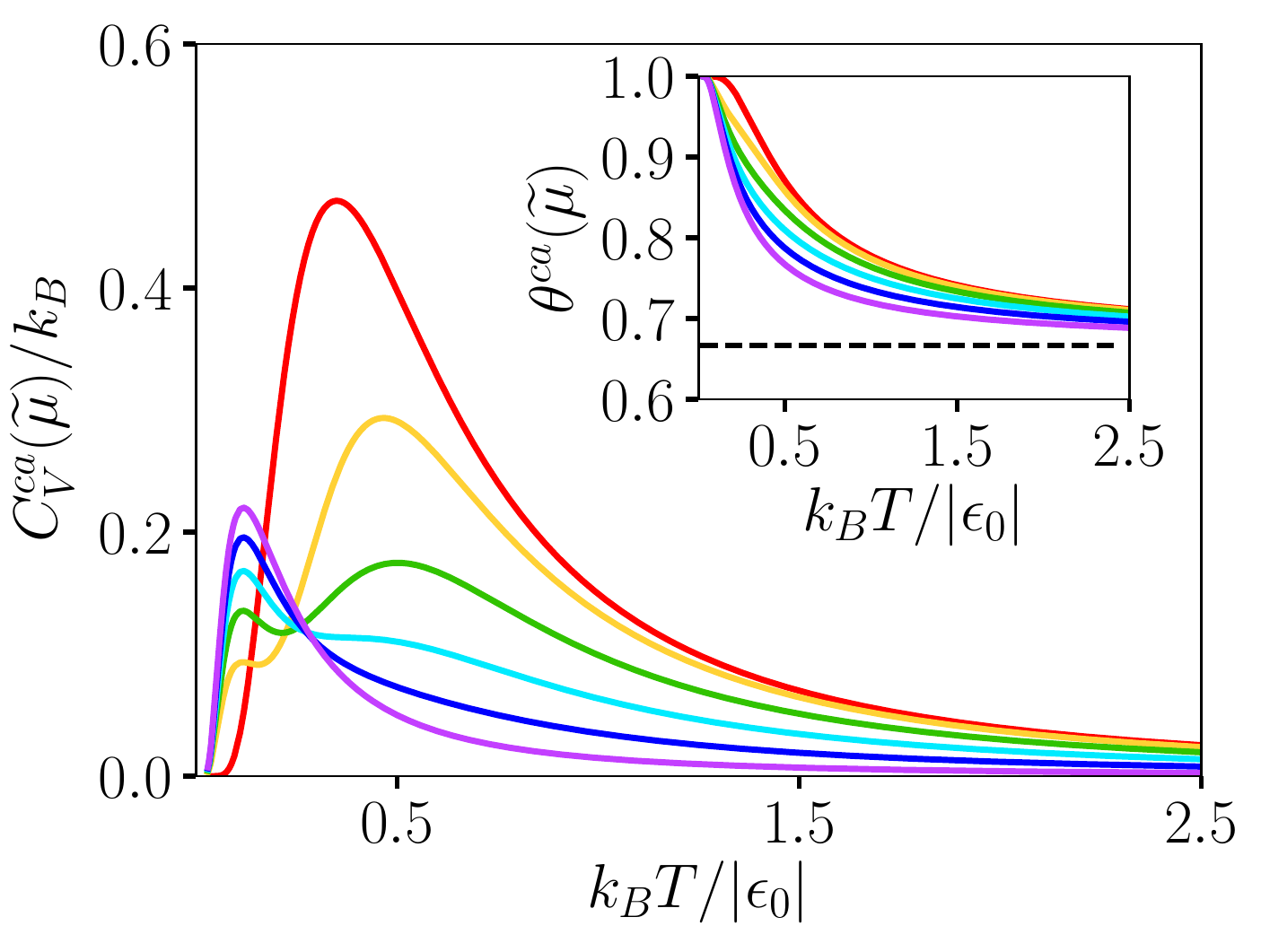}}
\subfigure[]{\includegraphics[scale=0.45]{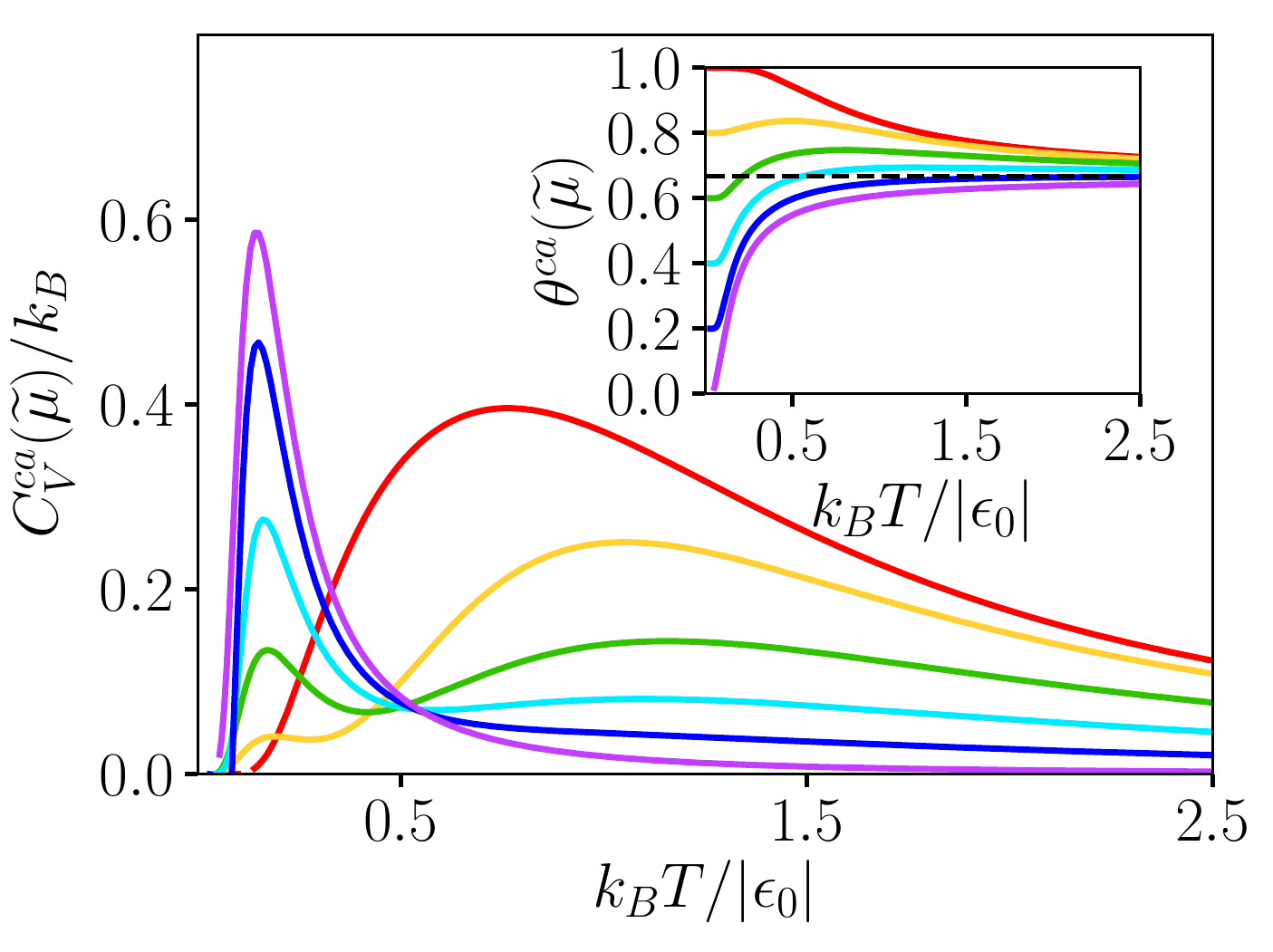}}
\caption{\label{consan2} (Color online) Specific heat for $c=c_{RW}=1.5$ for the same range of  probabilities and energy parameters as in Fig. \ref{consan} in the constrained annealing approach for the non-competitive scenario (a) $\Lambda = 2$  and competitive scenario  (b) $\Lambda = 2/3$. Insets: Corresponding helicity degrees where the dashed black line defines the asymptotic value of  eq. (\ref{23}) in the high temperature molten phase.}
\end{figure}

The fact that $\widetilde{\mu}$ is a function of $\beta\epsilon_0$ and $\beta\epsilon$ implies that we have  two possible scenarios depending on the value of 
\begin{eqnarray}
\Lambda \equiv |\epsilon_0/\epsilon|.
\end{eqnarray} 
If $\Lambda>1$ there is no competition between the favorable UA and unfavorable AA/UU pairs because in this case the sign of the energy of each pairing is determined only by $\epsilon_0$ and therefore is constant. We will call this the non-competitive regime. However when  $\Lambda<1$ the effect of quenched disorder becomes relevant since there is an effective energetic competition between different pairings  with a change of sign in the energy associated to favorable and unfavorable pairs. This case shall be referred to as the competitive regime. 
With $\epsilon_0<0$, although in both regimes $\epsilon_0 - \epsilon<0$, in the competitive regime $\epsilon_0 + \epsilon>0$  and $w^{ca}(\widetilde{\mu})$ exhibits a global maximum as well as a global minimum in the low temperature range independently from the specific values of $p$, $\epsilon_0$ and $\epsilon$, see Fig. \ref{weight}. 
As we will show next the presence of a global maximum in the statistical weight of base pairings is ultimately connected with the behavior of the helicity degree.  
When $\Lambda < 1$ the competition between favorable and unfavorable base pairs results in  the global maximum, for $0.5<p<1.0$, in $w(\widetilde{\mu})$, see fig, \ref{consan}.
From the comparison between specific heat and helicity with the respective constrained annealing weights for both  the competitive and non-competitive regimes it emerges that it is this competition that triggers the cold melting of the secondary structure which is described by the abrupt drop of the helicity at $T^*\approx  0.5|\epsilon_0|/k_B$ for different probabilities of U base occurrence $p$ with $c=c_{RW}=1.5$, see Fig.  \ref{consan2}.  
The first peak of the specific heat located in the range $0<T<T^*$ becomes more pronounced with respect to the case $\Lambda > 1$. Indeed each  specific heat peak corresponds to the gradual melting of the RNA secondary structure, the second of which is related to the usual hot melting.
This behavior is also reproduced  by setting $\epsilon_0=0$, i.e. by keeping a single degree of freedom in the Hamiltonian, although the specific heat in that scenario shows a single peak.
While for $0.5<p<1.0$ and $\epsilon_0 \ne 0$ the specific heat features always two peaks, also in the limiting cases $p=0.5$ and $p=1.0$, which correspond to annealed and homopolymeric case respectively, it exhibits a single peak in each of the two different temperature regions. 

\begin{figure*}[]
\centering
\subfigure[]{\includegraphics[scale=0.45]{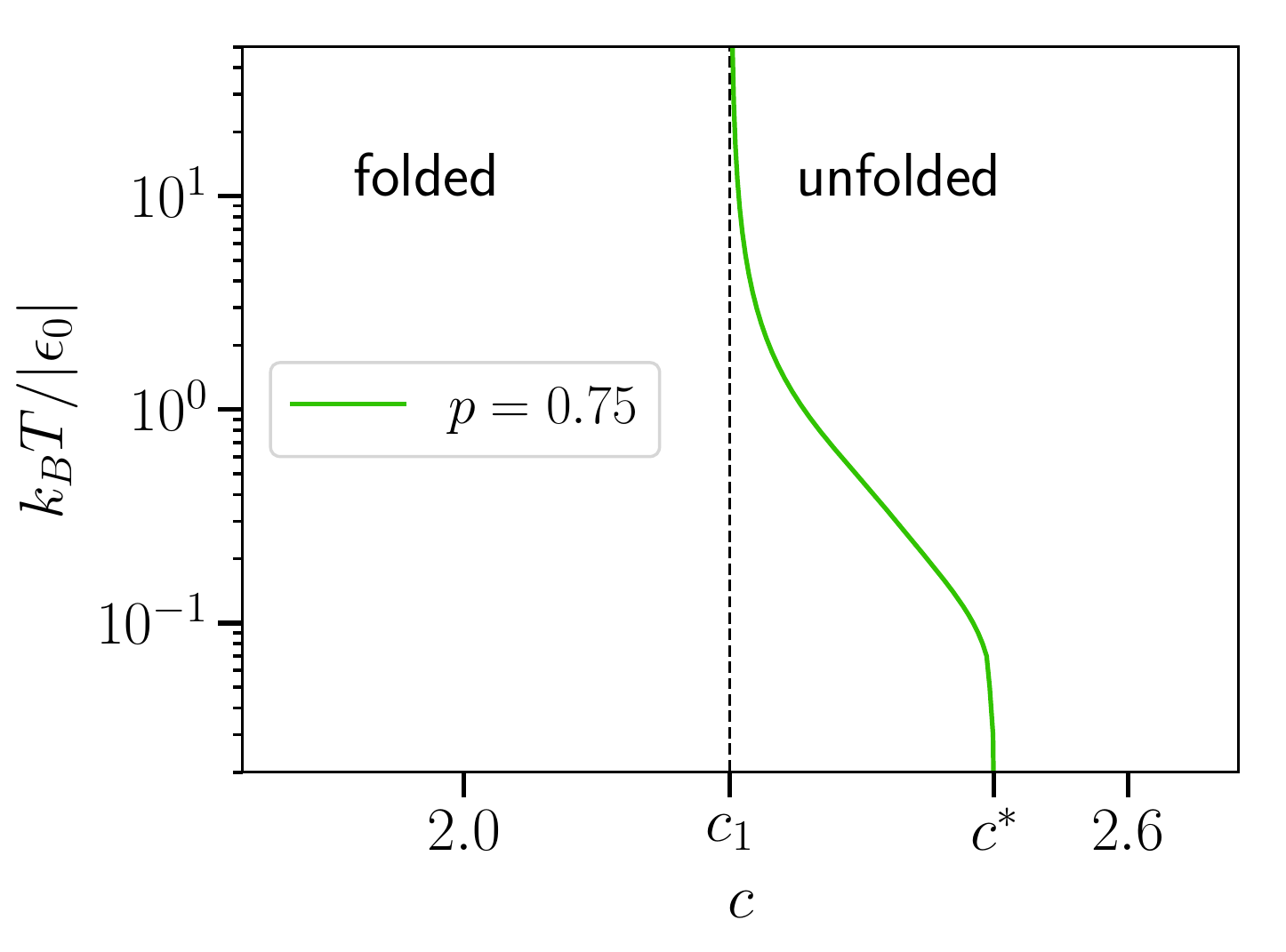}}
\subfigure[]{\includegraphics[scale=0.45]{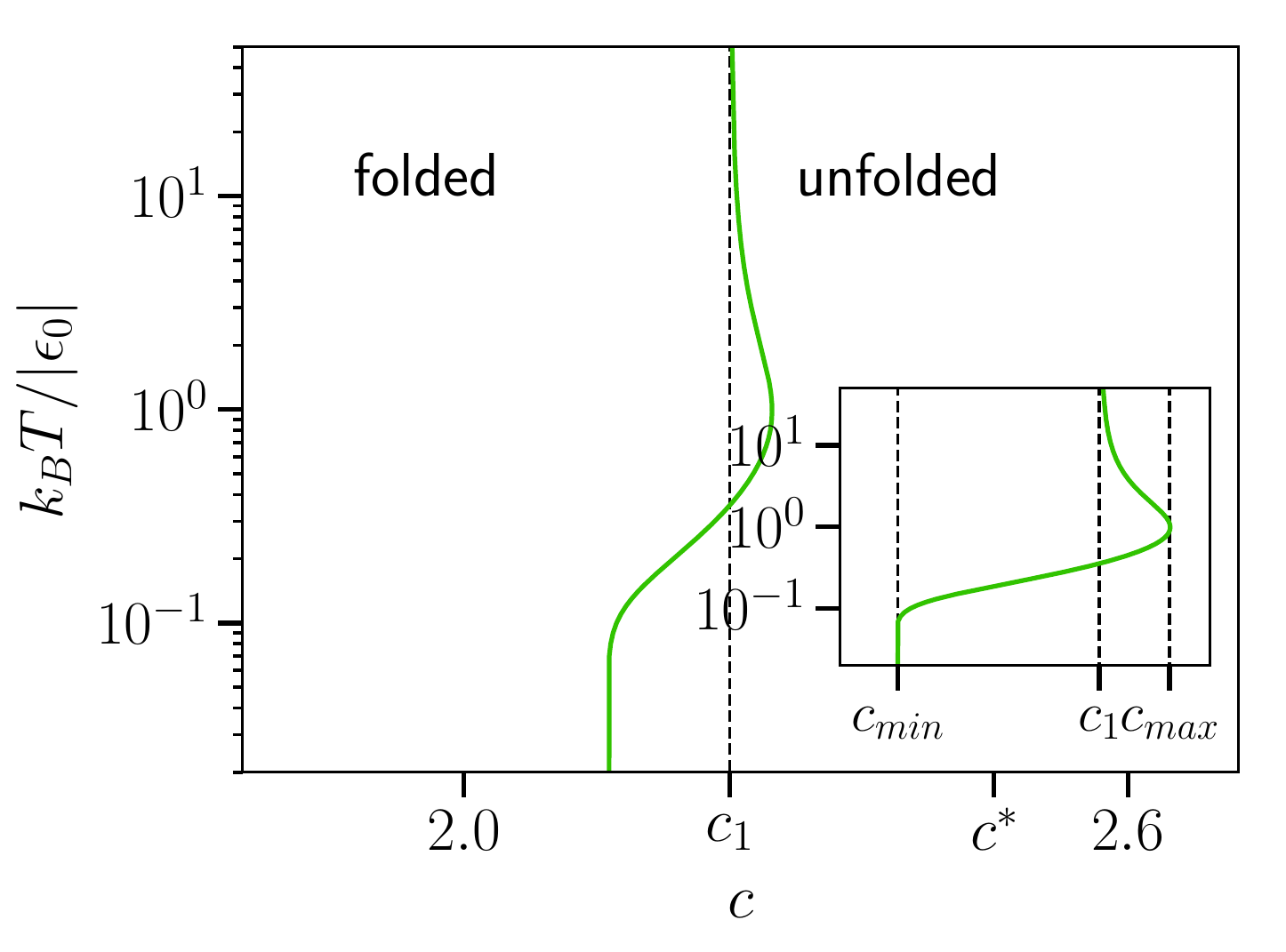}}\\
\subfigure[]{\includegraphics[scale=0.45]{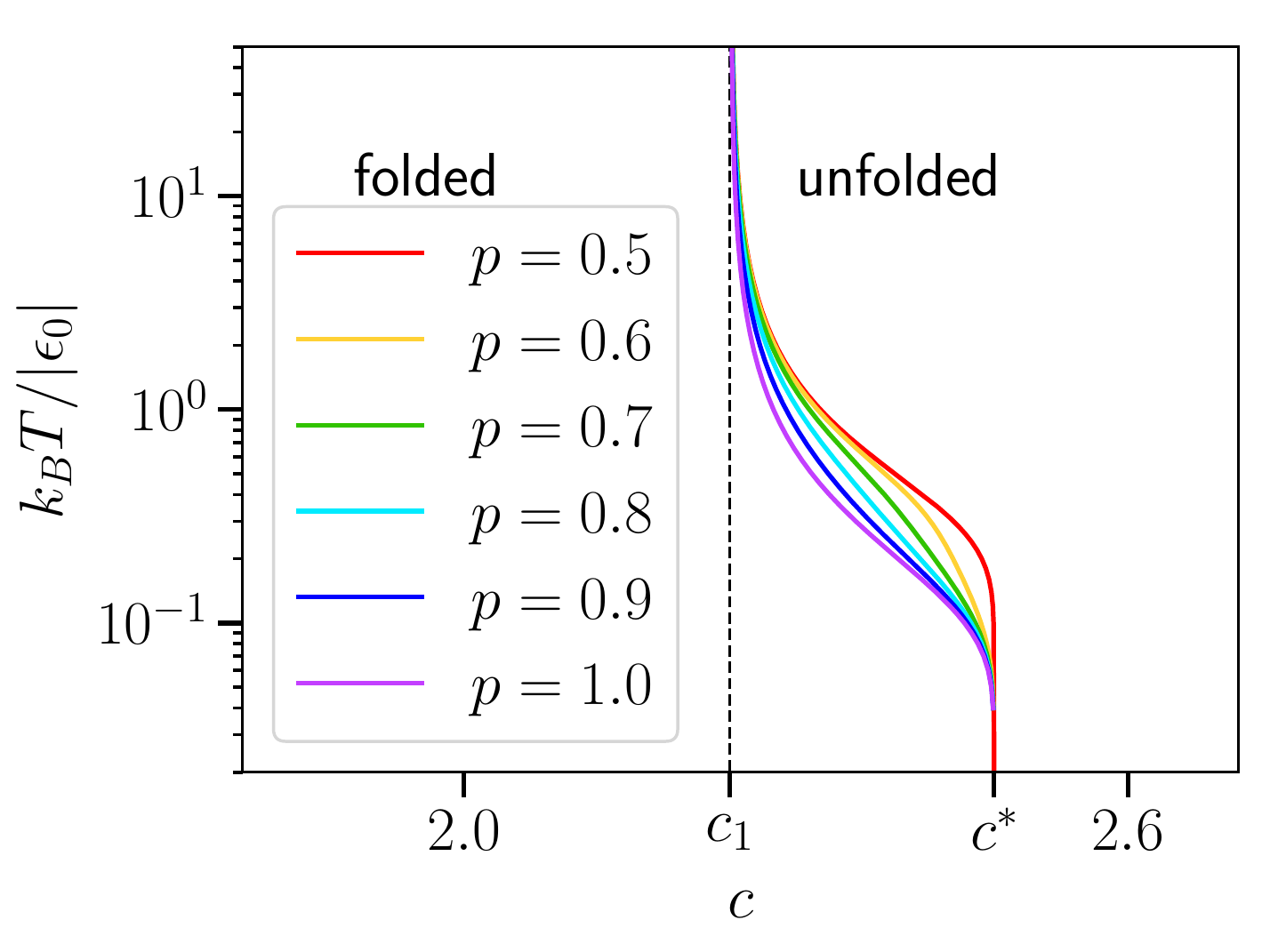}}
\subfigure[]{\includegraphics[scale=0.45]{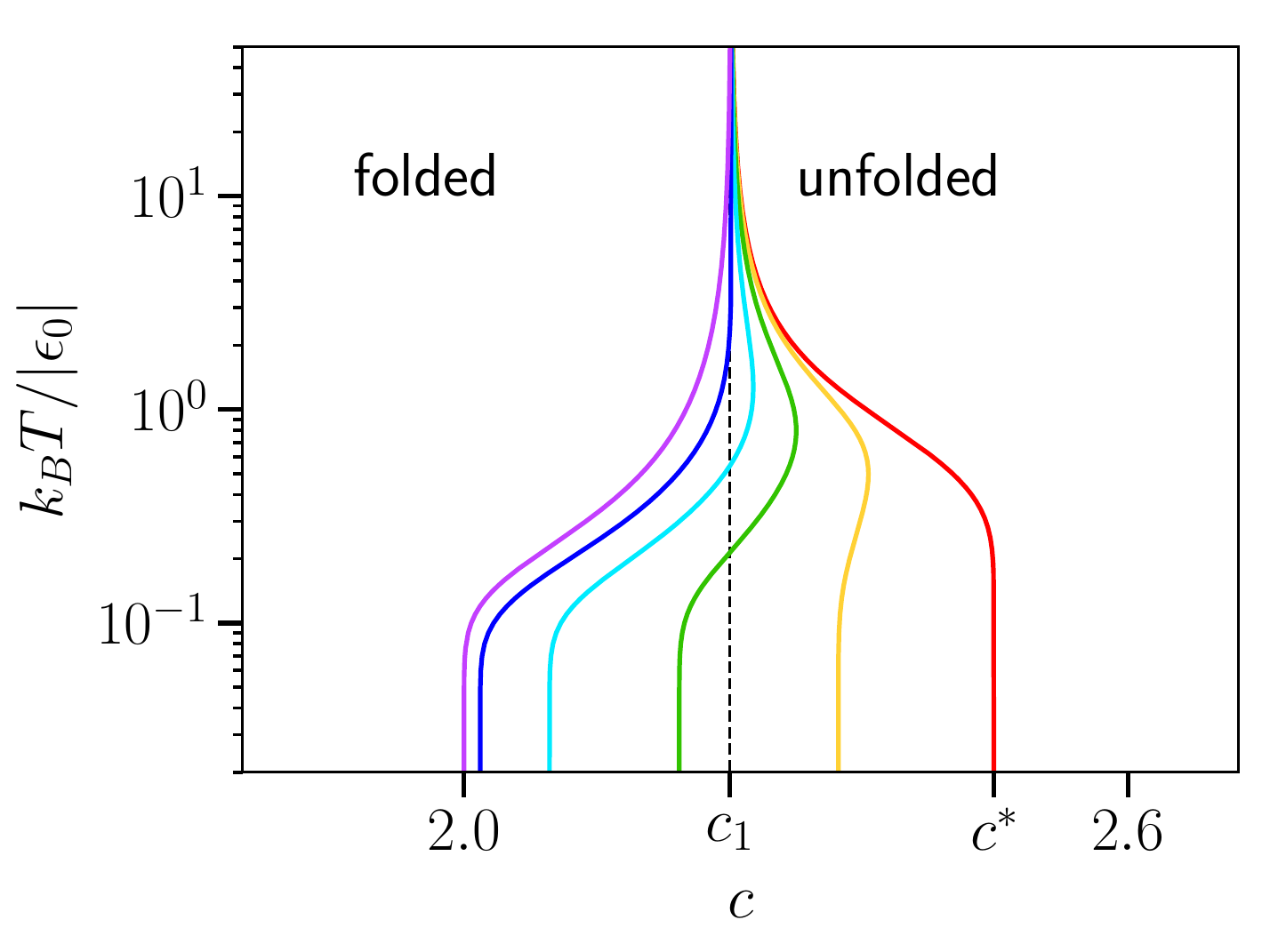}}
\caption{\label{phasea} (Color online) 
(a) Phase diagram for $\Lambda=2$ and $p=0.75$ in the $T-c$ plane. The critical line on which hot melting takes place is contained in the range $c_1<c<c^*$ of $w_m(c)$ so that for $c<c_1$ the molecule is always folded. (b) Phase diagram for $\Lambda=2/3$ and $p=0.75$ in the $T-c$ plane. For $c<c_{min}$ the molecule is always folded since $w_m(c)<w^{ca}(T)$, $\forall T$.  For $c_{min}<c<c_1$ there is only hot melting while  for $c>c_{max}$ the molecule is always unfolded since $w_m(c)>w^{ca}(T)$, $\forall T$. In the range $c_1<c<c_{max}$ the double intersection between $w_m(c)$ and $w^{ca}(T)$ determines also the onset of cold melting. Lower inset: Close up of the reentrant melting cross line. In (c) and (d) are shown the complete phase diagrams for $\Lambda=2$ and $\Lambda=2/3$ for the range of  probabilities from red to blue $p=0.5$, $p=0.6$, $p=0.7$, $p=0.8$, $p=0.9$ and $p=1.0$.}
\end{figure*}

\begin{figure*}
\centering
\subfigure[]{\includegraphics[scale=0.35]{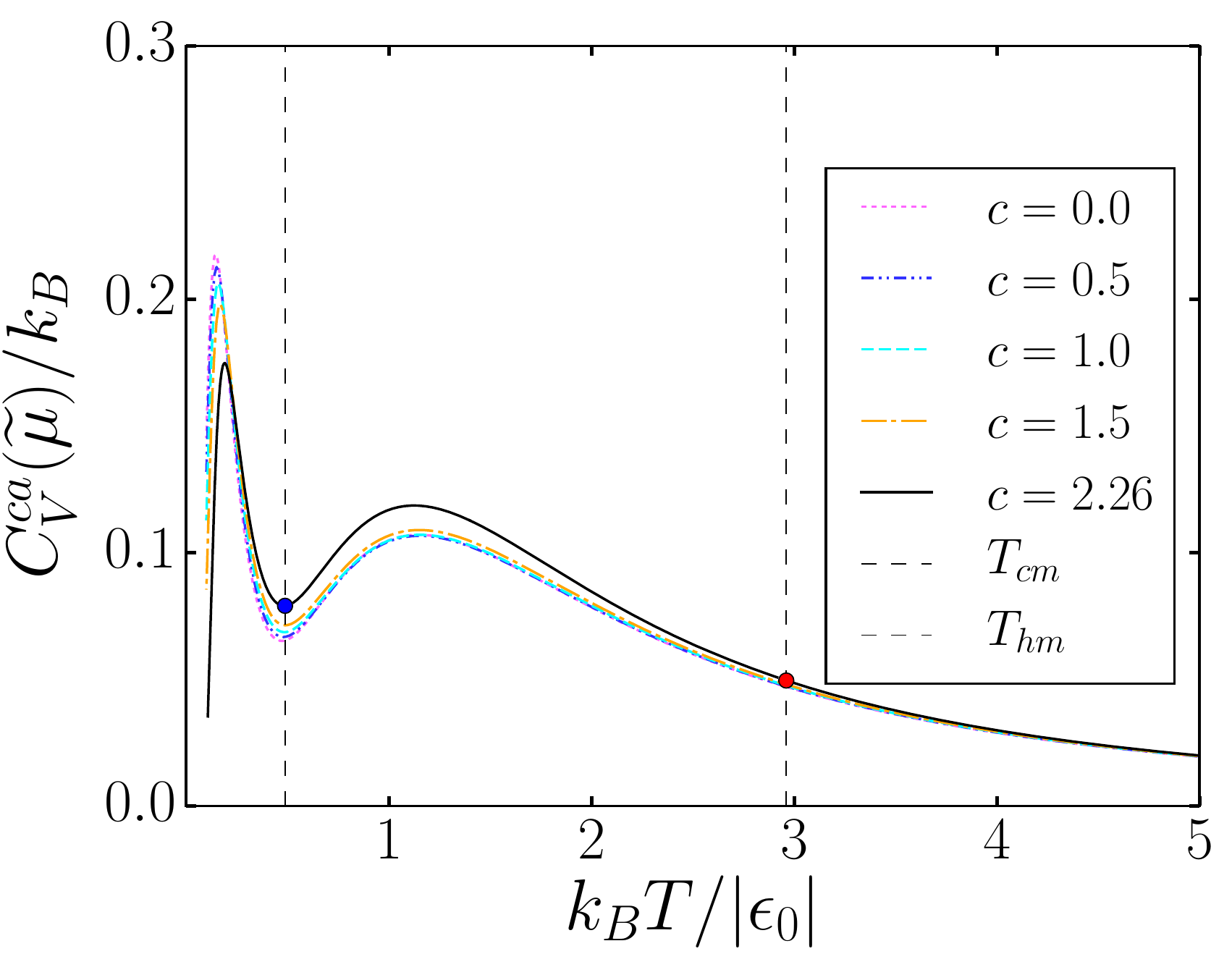}}
\subfigure[]{\includegraphics[scale=0.35]{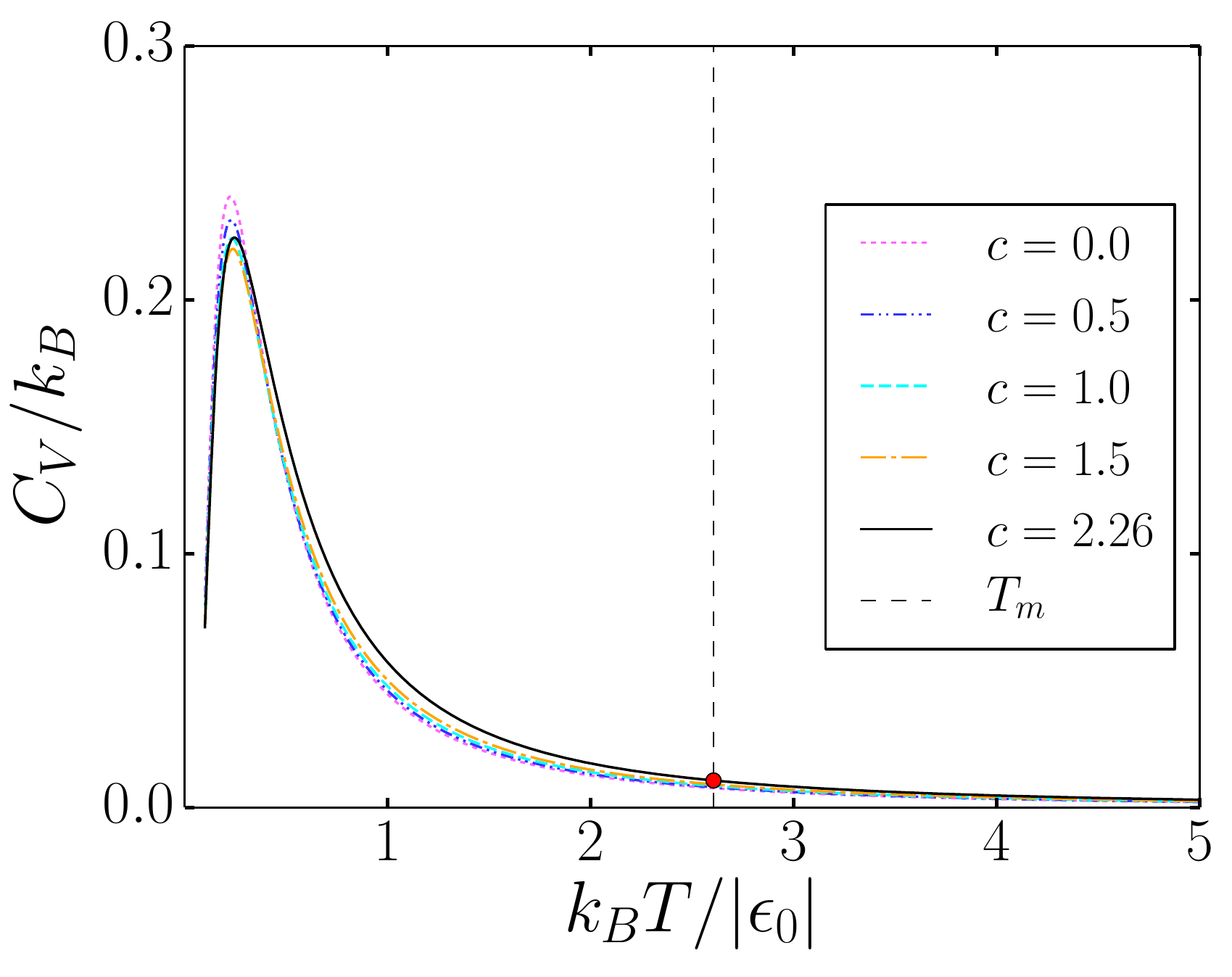}}
\subfigure[]{\includegraphics[scale=0.35]{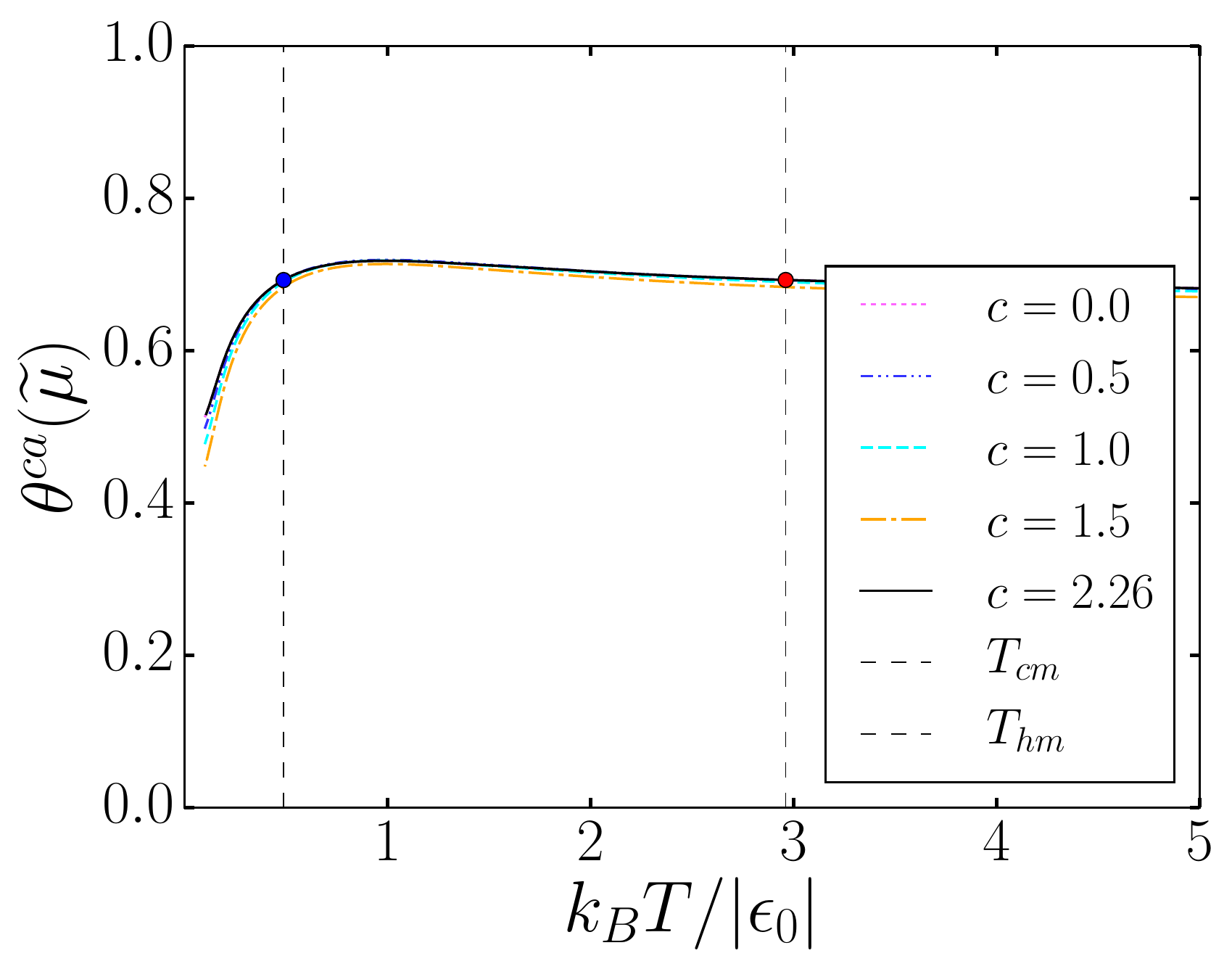}}
\subfigure[]{\includegraphics[scale=0.35]{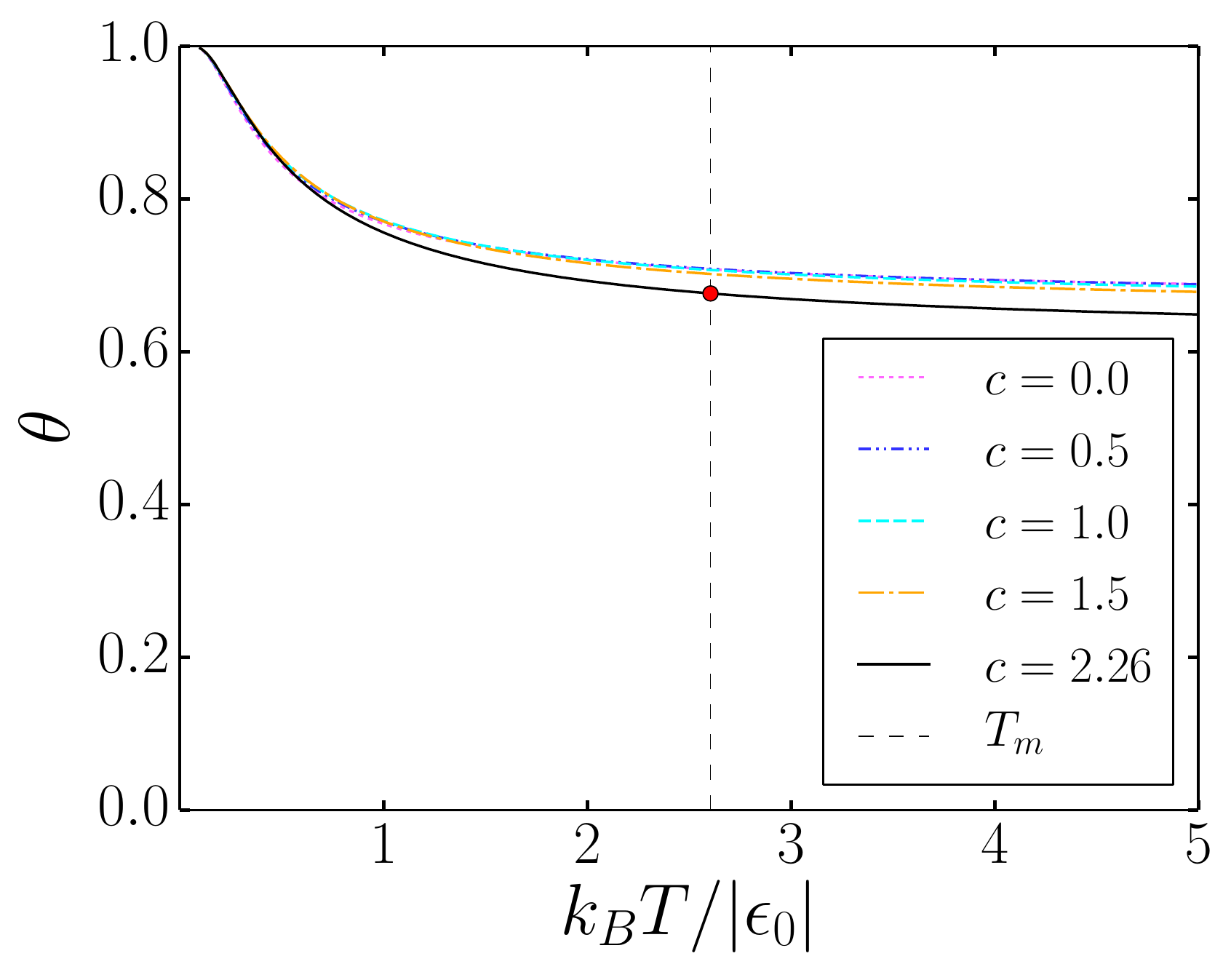}}
\caption{\label{heat}(Color online)  Constrained annealing specific heat in (a)  the disordered model for $p=0.75$ and $\epsilon = 1.5|\epsilon_0|$ resulting in $\Lambda = 2/3$ and  (b) the corresponding homopolymeric model for $\epsilon_0 < 0$  at various values of the loop exponent $c$ in the folded phase. In the disordered model the two critical temperatures $T_{cm}\approx 0.488|\epsilon_0|/k_B$ (blue circle) and $T_{hm} \approx 2.961|\epsilon_0|/k_B$ (red circle) correspond to the value of the loop exponent $c=2.26$ while for the homopolymer there is only $T_{m} \approx 2.605|\epsilon_0|/k_B $. In (c) and (d) is shown the corresponding  helicity. }
\end{figure*}

\subsection{Global phase diagrams}

In the competitive regime in addition to $c_1$, defined by eq.  (\ref{wc2_c1}), it is useful to define $c_{max}$ and $c_{min}$ respectively as 
\begin{eqnarray}
w_m(c_{max})\equiv\max_T w^{ca}(T),
\end{eqnarray}
\begin{eqnarray}
w_m(c_{min}) \equiv w^{ca}(T=0).
\end{eqnarray}
Then from the behaviour of $w^{ca}$ in the range $0.5<p<1.0$ there are only  four separate cases for the state of the molecule, depending on the value of the loop exponent. In fact there can be a double, single or no intersection at all between the values of $w^{ca}$ and $w_m(c)$ for each value of the loop exponent $c$.
If $p>p^*$, where $p^*$ is defined by  
\begin{eqnarray}
w_m(c_{min}(p^*)) =1,
\end{eqnarray}
the most interesting case appears for $c_1<c<c_{max}$, where $c_{max}$ depends on $p$, $\epsilon_0$ and $\epsilon$ while $c_1\approx 2.241$ is universal. Analogously  if $p<p^*$ one has $c_{min}(p) > c_1$ and the interesting range becomes $c_{min}<c<c_{max}$ as can be seen from the $p$-dependence of $w^{ca}$ in Fig. \ref{consan}.
In this scenario $w_m(c)$ intersects $w^{ca}(\widetilde{\mu})$  at two different temperatures $T_{cm}$ and $T_{hm}$. In addition to the high temperature melting also the cold melting found in \cite{roland1} takes place and manifests itself as a proper thermodynamic phase transition, where each phase is characterized by the relevant singularity of the homopolymeric grand-canonical partition function. By contrast in  the non-competitive regime, as for the homopolymer, only the hot melting transition takes place.
By fixing the background interaction $\epsilon_0<0$ and the probability at $p=0.75$, we solve the equation 
\begin{eqnarray}
w^{ca} (\widetilde{\mu}) = w_m(c),
\end{eqnarray}
which yields the values $\widetilde{\mu}_m (\beta \epsilon_0, \beta \epsilon, c)$ of the critical line in the phase diagram of  Fig. \ref{phasea} (a,b)  corresponding to $\Lambda = 2$ and $\Lambda = 2/3$ for $\epsilon = 0.5|\epsilon_0|$ and $\epsilon = 1.5|\epsilon_0|$ respectively. The reentrant melting point is defined by the crossing with  the asymptote $c=c_1$ in the case  $\Lambda < 1$ (b).
In figures \ref{phasea} (c) and (d) we show the global phase diagram for the complete range of the probabilities $0.5\le p \le 1.0$ in the competitive and non-competitive scenarios.

The c-dependence in the folded phase of the specific heat is very subtle for both the disordered model, where there is  a two peak structure, and the homopolymeric model where $C_V$ features only one peak, see Fig. \ref{heat} (a,b). A similar non significant spread is obtained for  the helicity behavior in Fig. \ref{heat} (c,d), where for the folded phase we use $\theta_b$ defined by eq. (\ref{thetab}) with $\kappa_b = \widetilde{\kappa_b^{ca}}$ and $z_b=z_b(\widetilde{\mu})$, while in the unfolded phase $\theta_p$ defined by eq.  (\ref{thetap}) with $w=w^{ca} (\widetilde{\mu})$.
The cold and the hot melting temperatures are estimated numerically for $c=2.26$, $p=0.75$, and $\epsilon = 1.5|\epsilon_0|$ as  $T_{cm}\approx 0.488|\epsilon_0|/k_B$ and $T_{hm}\approx 2.961|\epsilon_0|/k_B $ respectively.  
Quite remarkably the critical point for the reentrant transition lies almost exactly in the valley formed by the two peaks of the specific heat relative to the two different melting transitions.

\section{Conclusions}

In a two-letter RNA model without loop entropies, where no phase transition occurs, the heat capacities were shown to exhibit two peaks for the case of a quenched random sequence \cite{roland1}, indicating hot as well as cold melting.
For homopolymeric RNA, on the other hand, it is known that a finite loop entropy lead to a phase transition between a folded and an unfolded RNA state for a small range of the loop exponent $c$.  In the present paper we combine the main features of these two models and consider a two-letter RNA model with quenched randomness and in the presence of a finite loop entropy. As a main result, we show that for a small range of the loop exponent $c$ two phase transitions are encountered with changing the temperature, i.e. the folded state is only stable at intermediate temperatures. This might be related to the experimental observation of cold melting of RNA. With the present work by combining loop entropy with quenched randomness, which introduces energetic competition between different matching in the base pairs, we are able reproduce the cold denaturation phenomena and to describe it in the language of statistical mechanics and  phase transitions.     
Only for competing energies between favorable and unfavorable base pairs this transition occurs, as a result of the weakening of  the secondary structure formation due to quenched randomness in RNA sequences, as well as loop penalties which account for the existence of two relevant singularities in the grand-canonical partition function. 
The connection between the competitive regime and cold denaturation is investigated by comparing the constrained annealing weight and helicity with the specific heat  peaks in the low temperature range.
We argue that this transition is continuous and more specifically, since it is triggered by the same conformational effect of the homopolymeric counterpart, of order $n$, where $n$ is determined by  $(c-2)^{-1}-1<n<(c-2)^{-1}$ \cite{einert1}.
For the examined case with $c=2.26$ the reentrant melting transition is of third order and would become visible only by looking at higher order derivatives of the specific heat.
Finally, a particularly interesting direction for future works is to investigate the connection between cold denaturation and the glass transition for RNA molecules. 

\end{document}